\documentclass[review]{elsarticle}
\journal{Journal Of Computational Physics}

\def\ack{\section*{Acknowledgements}%
  \addtocontents{toc}{\protect\vspace{6pt}}%
  \addcontentsline{toc}{section}{Acknowledgements}%
}
\usepackage{lmodern}
\usepackage{lmodern}
\usepackage[T1]{fontenc}
\usepackage{amsmath}
\usepackage{amssymb}
\usepackage{graphicx}
\usepackage{esint}
\usepackage{mathtools}
\usepackage{color}
\usepackage{bm}
\biboptions{numbers,sort&compress}
\makeatletter

\usepackage{amsthm}\usepackage{latexsym}\usepackage{bm}\usepackage{amsfonts}

\usepackage{hyperref}

\usepackage{listings}
\usepackage{xcolor}
\begin{document}
\begin{frontmatter}
\title{High Order Explicit Lorentz Invariant Volume-preserving Algorithms
for Relativistic Dynamics of Charged Particles}

\author[address1]{Yulei Wang\corref{corresponding}}
\cortext[corresponding]{Corresponding author}
\ead{wyulei@nju.edu.cn}
\author[address2]{Jian Liu}
\author[address3]{Yang He}

\address[address1]{School of Astronomy and Space Science, Nanjing University,
Nanjing 210023, Peoples' Republic of China}
\address[address2]{Department of Engineering and Applied Physics, University of Science
and Technology of China, Hefei, 230026, Peoples' Republic of China}
\address[address3]{School of Mathematics and Physics, University of Science and Technology
Beijing, Beijing, 100083, Peoples' Republic of China}

\begin{abstract}
Lorentz invariant structure-preserving algorithms possess reference-independent
secular stability, which is vital for simulating relativistic multi-scale
dynamical processes. The splitting method has been widely used to
construct structure-preserving algorithms, but without exquisite considerations,
it can easily break the Lorentz invariance of continuous systems.
In this paper, we introduce a Lorentz invariant splitting technique
to construct high order explicit Lorentz invariant volume-preserving
algorithms (LIVPAs) for simulating charged particle dynamics. Using
this method, long-term stable explicit LIVPAs with different orders are developed
and their performances of Lorentz invariance and long-term stability
are analyzed in detail.
\end{abstract}
\begin{keyword}
\texttt{}Lorentz Invariant Volume-preserving Algorithms\sep Relativistic Charged Particles \sep High Order Explicit Schemes \sep Reference-independent Secular Stability
\end{keyword}

\end{frontmatter}

\section{Introduction\label{sec:Introduction}}

Structure-preserving algorithms possessing long-term stabilities play
key roles in simulations of various fields \cite{FengKang_1986,Forest_Ruth_1990,McLachlan_GeoAlgrithm_background,Candy_SympAlg_SepHam_1991,McLachlan_AccuracyOfsymInt1992,Cary_1993_VMPoison,ShangZaijiu_1999}.
In plasma studies, these advanced schemes have been applied in subjects
including the secular dynamical simulations of charged particles \cite{Qin_VariatianalSymlectic_2008,LiJinXing_GC_Symp_2011,Kraus_VariationalSym_Thesis},
the long-term analysis of plasma kinetic processes \cite{XiaoJY_PIC_wave_2015,CSPIC_2016,Kraus_VariationalSym_Thesis,QiangJi_2017_Sym_VMPoison,Shadwick_Variational_2014,Webb_2016_Sym_VMPoison},
the analysis of magnetohydrodynamical phenomena \cite{ZhouYao_2014MHD,ZhouYao_2016_PRE,XiaoJY_2016_NonCanSymTwoFluid},
and the simulations of nonlinear processes \cite{McLachlan_Symplectic_KDV,QinMengZhao_SympNSL}.
The basic idea of constructing structure-preserving algorithms is
to keep the fundamental geometry structures of ordinary differential
equations (ODEs) or partial differential equations (PDEs) during discretization \cite{McLachlan_GeoAlgrithm_background,Geometric_numerical_integration,Xiaojy_2018_PST}. In other words, the one-step maps, approximating the
continuous exact solutions, should inherit mathematical natures of
original continuous systems, such as the volume-preserving property
of source-free systems and the symplectic-preserving property of Hamiltonian
systems \cite{Geometric_numerical_integration}. These properties
work as constrains for numerical systems, which can limit the global
accumulations of numerical errors during long-term iterations \cite{Geometric_numerical_integration,Xiaojy_2018_PST}.

Explicit schemes, compared with implicit schemes, are more convenient
for implementations and more efficient especially in cases of complex
vector fields. Because the symplectic Runge-Kutta
methods are often implicit \cite{Geometric_numerical_integration},
the Hamiltonian splitting technique has been applied widely to build
explicit structure-preserving algorithms for charged particle dynamics.
Written in canonical coordinates, the Hamiltonian equation of charged
particles can be expressed as $J\dot{\mathbf{Z}}=\nabla_{\mathbf{Z}}H$,
where $\mathbf{Z}$ is the canonical coordinate, $J$ is the symplectic
structure, and $H$ is the Hamiltonian. Through the sum-split procedure
of $H$, one can obtain canonical symplectic subsystems which can
be discretized via standard symplectic methods, like the generating
function method \cite{ZhangRuili_ExpGenerateSym_2016,Zhangruili_2018}.
Then the composites of subsystem algorithms are the canonical symplectic
algorithm for the original system \cite{ZhouZhaoQi_2017_ExpSymp}.
On the other hand, it has been find that the Lorentz force equation
of charged particles possesses non-canonical symplectic structure,
which can be written as $K\dot{\mathbf{z}}=\nabla_{\mathbf{z}}H$,
where $\mathbf{z}$ is phase space coordinate and $K$ denotes the
$K$-symplectic structure. Similarly, the sum-split procedure of Hamiltonian
produces the explicit $K$-symplectic algorithms \cite{HeYang_Ksymp_PLA_2016,Xiaojy_2018_PST,XiaoJY_2019_NonSym}.
Compared with the symplectic methods, volume-preserving algorithms
(VPAs) impose looser constrains on discrete systems. However, VPAs
still possess significant long-term stability and have been used widely
in Particle-in-Cell simulations \cite{Birdsall_Book,PSC_2016,Ripperda_2018,Higuera_Cary_2017,VPA_covLorentz_2017}.
The volume-preserving systems can be expressed as $\dot{\mathbf{z}}=\mathbf{V}$,
where the source-free vector field $\mathbf{V}$ satisfies $\nabla_{\mathbf{z}}\cdot\mathbf{V}=0$.
Through splitting $\mathbf{V}$ into several source-free sub-vectors,
explicit VPAs can also be conveniently constructed \cite{Qin_Boris_2013,HeYang_Spliting_2015,Ruili_VPA_2015,He_2016_Highorder_RVPA,Higuera_Cary_2017,VPA_covLorentz_2017}.

For relativistic dynamical systems, the Lorentz invariance is another
fundamental property, which should also be kept after discretization.
When the effects of the special relativity are considered, simulating
a physical process in different Lorentz inertial frames can minimize
the range of time and space scales and thus reduce the cost of calculation
\cite{Vay_PRL_2007_RevEffets,Vay_2008}. In such cases, the Lorentz invariance
of algorithms becomes very important. In 2008, Vay has pointed out
the importance of Lorentz invariance for relativistic particles and
constructed a new scheme that preserves the $\mathbf{E}\times\mathbf{B}$
velocity in different frames \cite{Vay_2008}. In 2017, Higuera and
Cary improved Vay's method and built a VPA that also preserves $\mathbf{E}\times\mathbf{B}$
velocity \cite{Higuera_Cary_2017}. The reference-independency of the numerical
$\mathbf{E}\times\mathbf{B}$ velocity can be treated as one key outcome
of Lorentz invariance of algorithms. Strictly speaking, the Lorentz
invariant algorithms (LIAs) should produce reference-independent discrete
numerical results when solving the same process in arbitrary Lorentz
inertial frames. Equivalently, the difference equations of LIAs keep
unchanged after all observables are transformed to that in another
Lorentz inertial frame. Therefore, the definition of LIAs can be given
as follows \cite{Yulei_LCCSA_2016}. For a Lorentz invariant system
$\mathbf{F}$, an algorithm, $\mathcal{A}$, is Lorentz invariant
if it satisfies that
\begin{equation}
\mathcal{D}_{\mathcal{A}}\circ\mathcal{T}_{L}\mathbf{F}=\mathcal{T}_{L}\circ\mathcal{D}_{\mathcal{A}}\mathbf{F}\,\label{eq:LIADef}
\end{equation}
where, $\mathcal{D}_{\mathcal{A}}$ represents numerical discretization
operation using $\mathcal{A}$, $\mathcal{T}_{L}$ denotes the transformation
operation on variables based on the Lorentz matrix $L$, and \textquotedbl{}$\circ$\textquotedbl{}
is the composite operator. As an element of Lorentz group, $L$ satisfies
$LgL^{T}=g$, where $g$ is the Lorentz metric tensor \cite{Jackson_electrodynamics}.

The Lorentz invariance and the structure-preservation are two independent
aspects for constructions of advanced algorithms. Symplectic-preserving
or volume-preserving algorithms sometimes break the Lorentz invariance,
while traditional algorithms like the Newton
method and the Runge-Kutta method can produce Lorentz invariant schemes
if they are applied to Lorentz invariant dynamical
equations written in the 4-dimensional spacetime. Furthermore, if we combine the benefits of the structure-preservation
and the Lorentz invariance, the resulting structure-preserving LIAs will possess reference-independent
secular stability, which is very important for simulating relativistic
multi-scale processes. To construct Lorentz invariant algorithms,
one straightforward way is directly discretizing the Lorentz invariant
dynamical equations in which all variables are written as geometric objects in 4-dimensional spacetime \cite{Jackson_electrodynamics}.
The integrity of these geometric objects, such as the spacetime
coordinates and 4-dimensional tensors, should be kept during
discretization \cite{Yulei_LCCSA_2016}. For example, for the 4-dimensional
Lorentz invariant Hamiltonian equation of charged particles \cite{Jackson_electrodynamics},
the symplectic-Euler method gives an explicit 1-order symplectic LIA
\cite{Yulei_LCCSA_2016}. Similarly, symplectic Runge-Kutta methods,
such as the implicit mid-point symplectic scheme, also generate symplectic
LIAs when applied to this system.

Although symplectic LIAs with different orders can be constructed using
the symplectic Runge-Kutta methods, they are all implicit and their usage are not convenient.
On the other hand, although Hamiltonian splitting technique can produce explicit high-order
symplectic algorithms,
it has to divide the Hamiltonian into several pieces to construct symplectic schemes for subsystems \cite{ZhangRuili_ExpGenerateSym_2016,Zhangruili_2018,HeYang_Ksymp_PLA_2016,ZhouZhaoQi_2017_ExpSymp}. The fine-grained splitting breaks the Lorentz invariance. Therefore, in this paper, we
relax the constrain of symplectic-preservation and focus on the constructions
of high order explicit Lorentz invariant volume-preserving algorithms
(LIVPAs).
Higher order schemes mean faster convergence rate, but, unfortunately, more CPU time consumed for one-step iteration. Generally speaking, the balance between the complexity
and the convergence rate should be considered during practical simulations.
Though it's not necessary to build algorithms with arbitrary high orders, schemes with order higher than 1 are still needed. From the view point of applications,
the convergence rates of 1-order algorithms are sometimes too slow to
be applied. Therefore, although a little complex than 1-order algorithms,
2-order and 4-order schemes, like the Crank-Nicolson method (2-order),
the Boris method (2-order), and the 4-order Runge-Kutta method, have
been widely applied in plasma studies. Especially, for simulations
of processes with small scales, such as turbulence process and magnetic
reconnection processes, high temporal and spatial resolutions are
necessary. Though costing more CPU time for one-step iteration, high
order schemes, with larger step length, can be cheaper than low order
schemes.

Without exquisite considerations, the splitting technique can easily
break the Lorentz invariance of the original system, which will be
discussed in detail in Sec.\,\ref{sec:howsplitbreaklc}. In this
paper, however, we find a Lorentz invariant splitting procedure for
constructing LIVPAs for charged particle dynamics. The splitting method
is chosen for our kernel technique for two main reasons. First, through
different composite of subsystem algorithms, high-order explicit schemes
can be easily constructed \cite{Geometric_numerical_integration}.
Second, compared with the original systems, finding Lorentz invariant
algorithms for subsystems is much easier, and it is also readily to
understand that the composed scheme is Lorentz invariant if algorithms
of all subsystems are Lorentz invariant. Although we split the original
system into several pieces, the Lorentz invariance of algorithms still
remains. Explicit LIVPAs with different orders are constructed and
tested in detail. Compared with the explicit symplectic algorithms
constructed via Hamiltonian splitting method, these LIVPAs can give
reference-independent numerical results \cite{Zhangruili_2018}.
Compared with the Vay \cite{Vay_2008} and the Higuera-Cary \cite{Higuera_Cary_2017}
schemes that preserve the $\mathbf{E}\times\mathbf{B}$ velocity, LIVPAs show Lorentz invariance that is independent with the configurations of step length. Moreover, because of the preservation of phase space volume, LIVPAs show better secular stability than the Runge-Kutta method.

The rest part of this paper is organized as follows. In section \ref{sec:howsplitbreaklc},
basic properties of Lorentz invariant algorithms and their relations
with splitting method are discussed by use of a simple system. In
section \ref{sec:LorentzForceEquation}, the invariant form of the
Lorentz force equation is analyzed. We introduce the Lorentz invariant
splitting method and construct the LIVPAs in Sec.\,\ref{sec:constructLIVPAs}.
In section \ref{sec:NumericalExps}, LIVPAs are applied in a typical
field configuration and their numerical performances are studied.
Finally, we conclude this paper in Sec.\,\ref{sec:Conclusion}.

\section{LIAs and The Splitting Technique\label{sec:howsplitbreaklc}}

In this section, we give a general picture of LIAs. To simplify the expressions, here we use a simple 2-dimensional Lorentz invariant system $\mathbf{F}_2$,
\begin{equation}
\frac{\mathrm{d}\mathbf{x}}{\mathrm{d}\tau}=\mathbf{u}\,,\label{eq:introLIA2deq}
\end{equation}
where Lorentz invariant vectors $\mathbf{x}=\left(t,\,x\right)^T$ and $\mathbf{u}=\left(\gamma,\,p_x\right)^T$
denotes respectively the space-time coordinate and momentum. The proper time, $\tau$, is invariable under Lorentz transformations. $\mathbf{x}$ and $\mathbf{u}$ are functions of $\tau$.

Suppose there are two Lorentz inertial reference frames. One frame $\mathcal{O}$ is resting relative to lab, and another frame $\mathcal{O}'$ moves with constant speed $\beta$ relative to $\mathcal{O}$. In this case, the Lorentz matrix is
\begin{equation}
L_2=\frac{\partial \mathbf{x}'}{\partial \mathbf{x}}=\Gamma\left(\begin{array}{cc}
1 & -\beta \\
-\beta & 1 \\
\end{array}\right)\,,\label{eq:L_2}
\end{equation}
where $\Gamma=1/\sqrt{1-\beta^2}$, and the superscript "$'$" denotes physical
quantities observed in $\mathcal{O}'$. The Lorentz invariance of Equation \ref{eq:introLIA2deq}
means its form keeps unchanged after $\mathbf{x}$ and $\mathbf{u}$ are transformed into the moving
frame, namely, $\mathcal{T}_{L_2}\mathbf{F}_2$ gives
\begin{equation}
\frac{\mathrm{d}\mathbf{x}'}{\mathrm{d}\tau}=\mathbf{u}'\,.\label{eq:introLIA2deqMov}
\end{equation}

One way of discretization $\mathbf{F}_2$ is using the 1-order forward Euler difference $\mathcal{A}_1$, which does not break the integrity of $\mathbf{x}$ and $\mathbf{u}$. In frame $\mathcal{O}$, the difference equation, $\mathcal{D}_{\mathcal{A}_1}\mathbf{F}_2$, is
\begin{eqnarray}
\mathbf{x}^{k+1} & = & \mathbf{x}^{k}+\Delta\tau \mathbf{u}^k\,,\label{eq:introEuler1LIA}
\end{eqnarray}
where $k$ denotes "$k$th" step and $\Delta\tau$ is the step length. This algorithm can be proven to be Lorentz invariant as follows. First, replace $\mathbf{x}$ and $\mathbf{u}$ in Eq.\,\ref{eq:introEuler1LIA} with $L_2^{-1}\mathbf{x}'$ and $L_2^{-1}\mathbf{u}'$, respectively, which gives $L_2^{-1}\mathbf{x}'^{k+1}  =  L_2^{-1}\mathbf{x}'^{k}+\Delta\tau L_2^{-1}\mathbf{u}'^k$. Then, we can left-multiply $L_2$ on both side and obtain $\mathcal{T}_{L_2}\circ\mathcal{D}_{\mathcal{A}_1}\mathbf{F}_2$,
\begin{equation}
\mathbf{x'}^{k+1}  =  \mathbf{x}'^{k}+\Delta\tau \mathbf{u}'^k\,.\label{eq:introEuler1LIAprime}
\end{equation}
On the other hand, because the original system is Lorentz invariant, it is readily to see that the difference
equation $\mathcal{D}_{\mathcal{A}_1}\circ\mathcal{T}_{L_2}\mathbf{F}_2$ can be directly
obtained by discretizing Eq.\,\ref{eq:introLIA2deqMov} with algorithm $\mathcal{A}_1$, which has the
same form as Eq.\,\ref{eq:introEuler1LIAprime}. Consequently, for algorithm $\mathcal{A}_1$ and
system $\mathbf{F}_2$, we have
$\mathcal{T}_{L_2}\circ\mathcal{D}_{\mathcal{A}_1}\mathbf{F}_2=\mathcal{D}_{\mathcal{A}_1}\circ\mathcal{T}_{L_2}\mathbf{F}_2$.
Therefore, the 1-order forward Euler difference is Lorentz invariant.
Suppose that we set the initial condition as $\mathbf{x}^0$ in $\mathcal{O}$, the corresponding initial condition in $\mathcal{O}'$ is $\mathbf{x}'^0=L_2\mathbf{x}^0$. Because Eq.\,\ref{eq:introEuler1LIAprime} is directly
obtained from the Lorentz transformation of Eq.\,\ref{eq:introEuler1LIA}, numerical solutions $\mathbf{x}'^k$ of Eq.\,\ref{eq:introEuler1LIAprime} can also be calculated
from $\mathbf{x}^k$ given by Eq.\,\ref{eq:introEuler1LIA} via Lorentz transformation.
Therefore, for simulations of the same process, the difference equation of algorithm
$\mathcal{A}_1$ gives reference-independent numerical results in arbitrary Lorentz inertial frames.

To illustrate the effects of the splitting procedure on Lorentz invariance, we divide $\mathbf{u}$ into two parts, $\mathbf{u}=\mathbf{u}_1+\mathbf{u}_2=\left(\gamma,0\right)^T+\left(0,p_x\right)^T$.
Then we discretize the corresponding subsystems by different schemes, which is of common occurrence when constructing algorithms using splitting technique. We calculate $\mathbf{u}_1$ at $k$th step while $\mathbf{u}_2$ at $\left(k+1\right)$th step. After composing, the final algorithm  $\mathcal{D}_{\mathcal{A}_2}\mathbf{F}_2$ is
\begin{eqnarray}
t^{k+1}&=&t^{k}+\Delta\tau\gamma^{k}\,,\label{eq:S_1_S_2_tnlc}\\
x^{k+1}&=&x^{k}+\Delta\tau p_x^{k+1}\,.\label{eq:S_1_S_2_xnlc}
\end{eqnarray}

Now we derive $\mathcal{T}_{L}\circ\mathcal{D}_{\mathcal{A}_2}\mathbf{F}_2$. According to $\mathbf{x}=L_2^{-1}\mathbf{x}'$ and $\mathbf{u}=L_2^{-1}\mathbf{u}'$, we can directly substitute
$t=\Gamma\left(t'+\beta x'\right)$, $x=\Gamma\left(\beta t'+x'\right)$,
$\gamma=\Gamma\left(\gamma'+\beta p_x'\right)$, and $p_x=\Gamma\left(\beta \gamma'+ p_x'\right)$ into
Eqs.\,\ref{eq:S_1_S_2_tnlc} and \ref{eq:S_1_S_2_xnlc}. The resulting algorithm, $\mathcal{T}_{L_2}\circ\mathcal{D}_{\mathcal{A}_2}\mathbf{F}_2$, is
\begin{eqnarray}
t'^{k+1}&=&t'^{k}+\Delta\tau\Gamma^2\left(\gamma'^{k}-\beta^2\gamma'^{n+1}+\beta p'^k_x-\beta p'^{k+1}_x\right)\,,\label{eq:S_1_S_2_tnlcp}\\
x'^{k+1}&=&x'^{k}+\Delta\tau\Gamma^2\left(\beta\gamma'^{k+1}-\beta\gamma'^{n}+p'^{k+1}_x-\beta^2 p'^{k}_x\right)\,,\label{eq:S_1_S_2_tnlcp}
\end{eqnarray}
which is of different form compared with $\mathcal{D}_{\mathcal{A}_2}\circ\mathcal{T}_{L_2}\mathbf{F}_2$, namely,
\begin{eqnarray}
t'^{k+1}&=&t'^{k}+\Delta\tau\gamma'^{k}\,,\label{eq:S_1_S_2_tnlcp}\\
x'^{k+1}&=&x'^{k}+\Delta\tau p'^{k+1}_x\,.\label{eq:S_1_S_2_xnlcp}
\end{eqnarray}
Because $\mathcal{T}_{L_2}\circ\mathcal{D}_{\mathcal{A}_2}\mathbf{F}_2\neq\mathcal{D}_{\mathcal{A}_2}\circ\mathcal{T}_{L_2}\mathbf{F}_2$,
$\mathcal{A}_2$ is not Lorentz invariant. Even though we set the same initial condition, numerical results calculated by Eqs.\,\ref{eq:S_1_S_2_tnlcp}-\ref{eq:S_1_S_2_xnlcp} in frame $\mathcal{O}'$ cannot be
directly converted to results of Eqs.\,\ref{eq:S_1_S_2_tnlc}-\ref{eq:S_1_S_2_xnlc} via Lorentz transformation. In other words, for the
same process, the difference equation of algorithm $\mathcal{A}_2$ gives reference-dependent solutions
in different Lorentz inertial frames. One result of reference-dependency is the inconsistent numerical $\mathbf{E}\times\mathbf{B}$ velocity in different frames \cite{Vay_2008,Higuera_Cary_2017}.

Although the system discussed above is very simple and we won't use algorithms like $\mathcal{A}_2$ in practical simulations, we can clearly see the effects of the splitting method on the Lorentz invariance of original systems. The break of Lorentz invariant objects like $\mathbf{x}$ and $\mathbf{u}$ can, though not always, easily cause the violation of Lorentz invariance of numerical schemes.

\section{The Lorentz Force Equation of Charged Particles\label{sec:LorentzForceEquation}}

Before constructing the LIVPAs, we first analyze the Lorentz invariant form
of Lorentz force equation \cite{Jackson_electrodynamics}, namely,
\begin{eqnarray}
\frac{\mathrm{d}x^{\alpha}}{\mathrm{d}\tau} & = & U^{\alpha}\,,\nonumber \\
\frac{\mathrm{d}p^{\alpha}}{\mathrm{d}\tau} & = & \tilde{q}F^{\alpha\beta}U_{\beta}\,,\label{eq:CovLorentzForce_XP}
\end{eqnarray}
where $x^{\alpha}=\left(t,x,y,z\right)^{T}$ is the coordinate in
4-dimensional spacetime, $U^{\alpha}=\left(\gamma,\gamma v_{x},\gamma v_{y},\gamma v_{z}\right)^{T}$
is the 4-velocity, $\gamma=\sqrt{1+p^{2}}$ is the Lorentz
factor, $p^{\alpha}=\left(\gamma,p_{x},p_{y},p_{z}\right)^{T}$ is
the 4-momentum, and
\begin{equation}
F^{\alpha\beta}=\left(\begin{array}{cccc}
0 & -E_{x} & -E_{y} & -E_{z}\\
E_{x} & 0 & -B_{z} & B_{y}\\
E_{y} & B_{z} & 0 & -B_{x}\\
E_{z} & -B_{y} & B_{x} & 0
\end{array}\right)\label{eq:F}
\end{equation}
is the electromagnetic tensor which is the function of $x^{\alpha}$.
$E_{x}$, $E_{y}$, $E_{z}$, $B_{x}$, \textbf{$B_{y}$}, and\textbf{
}$B_{z}$ denote the space components of the electric field and magnetic
field, respectively. $\tilde{q}$ denotes the charge sign, namely,
$\tilde{q}=1$ for positive charged particles and $\tilde{q}=-1$
for negative charged particles. In this paper, without special instructions,
all physical quantities are normalized according to the units given
in Tab.\,\ref{tab:Units} and the space components of vectors and
tensors are written in the Cartesian coordinate system. The superscripts
and subscripts written in Greek alphabets denote "contravariant" and
"covariant" components respectively. They can convert to each other
through the Lorentz metric tensor $g_{\alpha\beta}$ satisfying
\begin{equation}
g^{\alpha\beta}=g_{\alpha\beta}=\left(\begin{array}{cccc}
1 & 0 & 0 & 0\\
0 & -1 & 0 & 0\\
0 & 0 & -1 & 0\\
0 & 0 & 0 & -1
\end{array}\right)\,.\label{eq:g}
\end{equation}
For example, $x_{\alpha}=g_{\alpha\beta}x^{\beta}=\left(t,-x,-y,-z\right)^{T}$.
Einstein's convention for the summation over repeat indices is applied
in this paper.

\begin{table}[h]
\centering
\begin{tabular}{ccc}
\hline
\textbf{Physical quantities} & \textbf{Symbols} & \textbf{Units}\tabularnewline
\hline
Time & $t$,$\tau$ & $\mathrm{m_{0}}/qB_{0}$\tabularnewline
Space & $x^{\alpha}$ & $\mathrm{m_{0}c}/qB_{0}$\tabularnewline
Momentum & $p^{\alpha}$ & $\mathrm{m_{0}c}$\tabularnewline
Velocity & $U^{\alpha}$, $\bm{\beta}$ & $\mathrm{c}$\tabularnewline
Electric strength & $\mathbf{E}$ & $B_{0}\mathrm{c}$\tabularnewline
Magnetic strength & $\mathbf{B}$ & $B_{0}$\tabularnewline
Vector field & $\mathbf{A}$ & $\mathrm{m}_{0}\mathrm{c}/q$\tabularnewline
Scalar field & $\phi$ & $\mathrm{m}_{0}\mathrm{c}^{2}/q$\tabularnewline
Energy & $\mathcal{H}$, $H$, $\gamma$ & $\mathrm{m_{0}c^{2}}$\tabularnewline
\hline
\end{tabular}

\caption{Units of all the physical quantities used in this paper. $\mathrm{m}_{0}$
is the rest mass of a particle, $q$ is the norm of charge carried
by a particle, $\mathrm{c}$ is the speed of light, and $B_{0}$ is
the reference strength of magnetic field.\label{tab:Units}}
\end{table}

Now we organize Eq.\,\ref{eq:CovLorentzForce_XP} into a form of
ordinary differential equations in terms of $x^{\alpha}$ and $p^{\alpha}$.
Considering that the normalized 4-velocity and 4-momentum have the
same value, namely, $p^{\alpha}=U^{\alpha}=\left(\gamma,p_{x},p_{y},p_{z}\right)^{T}$.
Eq.\,\ref{eq:CovLorentzForce_XP} can thus be rewritten as
\begin{eqnarray}
\frac{\mathrm{d}x^{\alpha}}{\mathrm{d}\tau} & = & p^{\alpha}\,,\nonumber \\
\frac{\mathrm{d}p^{\alpha}}{\mathrm{d}\tau} & = & \tilde{q}F_{\ \beta}^{\alpha}p^{\beta}\,,\label{eq:CovLorentzForceM_XP}
\end{eqnarray}
where
\begin{equation}
F_{\ \beta}^{\alpha}=F^{\alpha\xi}g_{\xi\beta}=\left(\begin{array}{cccc}
0 & E_{x} & E_{y} & E_{z}\\
E_{x} & 0 & B_{z} & -B_{y}\\
E_{y} & -B_{z} & 0 & B_{x}\\
E_{z} & B_{y} & -B_{x} & 0
\end{array}\right)\,.\label{eq:Fa_b}
\end{equation}
Defining $\mathcal{Z}=\left(x^{\alpha},p^{\alpha}\right)^{T}$, we can transform Eq.\,\ref{eq:CovLorentzForceM_XP} to a compact form
\begin{equation}
\dot{\mathcal{Z}}=\mathcal{V}\left(\mathcal{Z}\right)\,,\label{eq:dZ}
\end{equation}
where the dot operator denotes the full derivative $\mathrm{d}/\mathrm{d}\tau$,
and the vector field on right side $\mathcal{V}\left(\mathcal{Z}\right)=\left(p^{\alpha},\tilde{q}F_{\ \beta}^{\alpha}p^{\beta}\right)^{T}$can
also be represented by the Lie derivative
\begin{equation}
X_{\mathcal{V}}=p^{\alpha}\frac{\partial}{\partial x^{\alpha}}+\tilde{q}F_{\ \beta}^{\alpha}p^{\beta}\frac{\partial}{\partial p^{\alpha}}\,.\label{eq:LieXv}
\end{equation}
Therefore, Eq.\,\ref{eq:dZ} becomes $\dot{\mathcal{Z}}=X_{\mathcal{V}}\mathcal{Z}$,
whose analytical solution can be generated by the one-parameter Lie
group $\exp\left(\tau X_{\mathcal{V}}\right)$ as $\phi^{\tau}\coloneqq\exp\left(\tau X_{\mathcal{V}}\right)\mathcal{Z}\left(0\right)$
\cite{Ruili_VPA_CiCP_2016}. Considering that
\begin{equation}
\nabla_{\mathcal{Z}}\cdot\mathcal{V}=\frac{\partial p^{\alpha}}{\partial x^{\alpha}}+\frac{\partial}{\partial p^{\alpha}}\left(\tilde{q}F_{\ \beta}^{\alpha}p^{\beta}\right)=0\,,\label{eq:sourcfree}
\end{equation}
$\mathcal{V}$ is a source-free vector field, and, the solution $\phi^{\tau}$
of Eq.\,\ref{eq:CovLorentzForceM_XP} is a volume-preserving map.

\section{Constructions of LIVPAs\label{sec:constructLIVPAs}}

In this section, we construct explicit high-order LIVPAs of Eq.\,\ref{eq:CovLorentzForceM_XP}
by using splitting technique. To keep the Lorentz invariance of
subsystems, we want to keep the integrity of invariant objects, $x^{\alpha}$, $p^{\alpha}$,
and $F_{\ \beta}^{\alpha}$, and the first choice of splitting $\mathcal{V}\left(\mathcal{Z}\right)$ seems to be
\begin{equation}
\mathcal{V}\left(\mathcal{Z}\right)=\mathcal{V}_{p}+\mathcal{V}_{F}=\left(\begin{array}{c}
p^{\alpha}\\
0
\end{array}\right)+\left(\begin{array}{c}
0\\
\tilde{q}F_{\ \beta}^{\alpha}p^{\beta}
\end{array}\right)\,.\label{eq:VzSplitPF}
\end{equation}
However, for arbitrary $F_{\ \beta}^{\alpha}$, it's hard to find
the exact solution or volume-preserving approximations for the subsystem
of $\mathcal{V}_{F}$. Therefore, we have to seek other proper ways
of splitting $F_{\ \beta}^{\alpha}$, which should satisfy that
\begin{enumerate}
\item the subsystems can be solved by explicit volume-preserving numerical
schemes;
\item the algorithms of subsystems should be Lorentz invariant.
\end{enumerate}
In the following three subsections, we first exhibit the Lorentz invariant splitting
procedure which generates volume-preserving Lorentz invariant subsystems. Second, we construct
the LIVPAs for subsystems. Third, we construct the final LIVPAs with different orders.

\subsection{The Lorentz invariant volume-preserving splitting procedure}
In this section, three Lorentz inertial frames, $\mathcal{O}^r$, $\mathcal{O}$, and $\mathcal{O}'$, will be involved. To give clear expressions, here we describe the definitions of the three frames and several symbols that will be frequently used. We call $\mathcal{O}^{r}$ the splitting reference frame (SRF), which works as a medium linking the numerical results in arbitrary Lorentz inertial reference frames through Lorentz transformation. $\mathcal{O}$ and $\mathcal{O}'$ are two arbitrary Lorentz inertial frames. Symbols of main observables we use in the three frames are listed in Tab.\,\ref{tab:Osymbol}. And the definitions of three Lorentz transformation matrices, $L$, $M$, and $N$, are listed in Tab.\,\ref{tab:LMN}. In $\mathcal{O}$, the electromagnetic tensor is $F_{\ \beta}^{\alpha}$, and, if observed in frame $\mathcal{O}^{r}$ (transformed back to frame $\mathcal{O}^r)$, it becomes \cite{Jackson_electrodynamics}
\begin{equation}
\mathcal{F}_{\ \beta}^{\alpha}=L^{-1}F_{\ \beta}^{\alpha}L=\left(\begin{array}{cccc}
0 & E_{x}^{r} & E_{y}^{r} & E_{z}^{r}\\
E_{x}^{r} & 0 & B_{z}^{r} & -B_{y}^{r}\\
E_{y}^{r} & -B_{z}^{r} & 0 & B_{x}^{r}\\
E_{z}^{r} & B_{y}^{r} & -B_{x}^{r} & 0
\end{array}\right)\,.\label{eq:F_rest}
\end{equation}
In $\mathcal{O}^r$, the kinetic tensor $\mathcal{K}_{\ \beta}^{\alpha}$ and the rotation tensor $\mathcal{R}_{\ \beta}^{\alpha}$
are respectively the electric and magnetic parts of $\mathcal{F}_{\ \beta}^{\alpha}$,
namely,
\begin{eqnarray}
\mathcal{K}_{\ \beta}^{\alpha} & = & \left(\begin{array}{cccc}
0 & E_{x}^{r} & E_{y}^{r} & E_{z}^{r}\\
E_{x}^{r} & 0 & 0 & 0\\
E_{y}^{r} & 0 & 0 & 0\\
E_{z}^{r} & 0 & 0 & 0
\end{array}\right)\,,\label{eq:scriptK_def}\\
\mathcal{R}_{\ \beta}^{\alpha} & = & \left(\begin{array}{cccc}
0 & 0 & 0 & 0\\
0 & 0 & B_{z}^{r} & -B_{y}^{r}\\
0 & -B_{z}^{r} & 0 & B_{x}^{r}\\
0 & B_{y}^{r} & -B_{x}^{r} & 0
\end{array}\right)\,.\label{eq:scriptR_def}
\end{eqnarray}
$\mathcal{K}_{\ \beta}^{\alpha}$ corresponds to the change of kinetic
energy, while the $\mathcal{R}_{\ \beta}^{\alpha}$ part corresponds
to the rotation of momentum.

\begin{table}[h]
\centering
\begin{tabular}{cccc}
\hline
\textbf{Observables} & in $\mathcal{O}^r$ & in $\mathcal{O}$ & in $\mathcal{O}'$\tabularnewline
\hline
Coordinate & $x^{r,\alpha}$ & $x^\alpha$ & $x'^{\alpha}$\tabularnewline
Momentum & $p^{r,\alpha}$ & $p^\alpha$ & $p'^{\alpha}$\tabularnewline
Electric Field & $\mathbf{E}^{r}$ & $\mathbf{E}$& $\mathbf{E}'$\tabularnewline
magnetic Field & $\mathbf{B}^{r}$ & $\mathbf{B}$& $\mathbf{B}'$\tabularnewline
Electromagnetic tensor & $\mathcal{F}^{\alpha}_{\ \beta}$ & $F^{\alpha}_{\ \beta}$& $F_{\ \beta}'^{\alpha}$\tabularnewline
Kinetic tensor & $\mathcal{K}^{\alpha}_{\ \beta}$ & $K^{\alpha}_{\ \beta}$& $K_{\ \beta}'^{\alpha}$\tabularnewline
Rotation tensor & $\mathcal{R}^{\alpha}_{\ \beta}$ & $R^{\alpha}_{\ \beta}$& $R_{\ \beta}'^{\alpha}$\tabularnewline
Square of Kinetic tensor & $\mathcal{S}^{\alpha}_{\ \beta}$ & $S^{\alpha}_{\ \beta}$& $S_{\ \beta}'^{\alpha}$\tabularnewline
Square of Rotation tensor & $\mathcal{P}^{\alpha}_{\ \beta}$ & $P^{\alpha}_{\ \beta}$& $P_{\ \beta}'^{\alpha}$\tabularnewline
\hline
\end{tabular}

\caption{List of observable symbols in $\mathcal{O}^r$, $\mathcal{O}$, and $\mathcal{O}'$. The detailed definitions of kinetic and rotation tensors are given by Eqs.\,\ref{eq:scriptK_def} and \ref{eq:scriptR_def}. To simplify the expressions, we defines  $\mathcal{S}_{\ \beta}^{\alpha}=\left(\mathcal{K}_{\ \beta}^{\alpha}\right)^{2}$ and
$\mathcal{P}_{\ \beta}^{\alpha}=\left(\mathcal{R}_{\ \beta}^{\alpha}\right)^{2}$.\label{tab:Osymbol}}
\end{table}

\begin{table}[h]
\centering
\begin{tabular}{ccc}
\hline
\textbf{Lorentz matrix} & \textbf{Definition} & \textbf{Description} \tabularnewline
\hline
$L$ & $\partial x^\alpha/x^{r,\beta}$ & $\mathcal{O}$ relative to $\mathcal{O}^r$\tabularnewline
$M$ & $\partial x'^\alpha/x^{\beta}$ & $\mathcal{O}'$ relative to $\mathcal{O}$\tabularnewline
$N$ & $ML=\partial x'^\alpha/x^{r,\beta}$ &$\mathcal{O}'$ relative to $\mathcal{O}^r$\tabularnewline
\hline
\end{tabular}

\caption{Definitions of the Lorentz transformation matrices $L$, $M$, and $N$. To simplify the expressions, in this paper, we do not explicitly write the upper and lower indexes of $L$, $M$, and $N$.\label{tab:LMN}}
\end{table}

Now we introduce the splitting procedure of the original system. In frame $\mathcal{O}$, we split the vector field $\mathcal{V}$ into three
parts, namely,
\begin{equation}
\mathcal{V}\left(\mathcal{Z}\right)=\left(\begin{array}{c}
p^{\alpha}\\
\tilde{q}F_{\ \beta}^{\alpha}p^{\beta}
\end{array}\right)=\mathcal{V}_{p}+\mathcal{V}_{K}+\mathcal{V}_{R}=\left(\begin{array}{c}
p^{\alpha}\\
0
\end{array}\right)+\left(\begin{array}{c}
0\\
\tilde{q}K_{\ \beta}^{\alpha}p^{\beta}
\end{array}\right)+\left(\begin{array}{c}
0\\
\tilde{q}R_{\ \beta}^{\alpha}p^{\beta}
\end{array}\right)\,,\label{eq:VsplitLIA}
\end{equation}
where $K_{\ \beta}^{\alpha}=L\mathcal{K}_{\ \beta}^{\alpha}L^{-1}$,
$R_{\ \beta}^{\alpha}=L\mathcal{R}_{\ \beta}^{\alpha}L^{-1}$. There are three main reasons for splitting
$\mathcal{V}$ in this way. First, $\mathcal{V}_{p}$, $\mathcal{V}_K$, and $\mathcal{V}_R$
are source-free vector fields, which generate volume-preserving subsystems.
Second, as will be shown later, because $K_{\ \beta}^{\alpha}$ and
$R_{\ \beta}^{\alpha}$ are related to the kinetic and rotation tensors observed in
the SRF $\mathcal{O}^{r}$, the Lorentz invariance of
the subsystems and the corresponding algorithms can be easily kept.
Third, such splitting technique can significantly simplify the expressions
of final difference equations.

The three parts of $\mathcal{V}$ can be proven to be source free
vector fields as follows. For $\mathcal{V}_{p}$, $\nabla_{\mathcal{Z}}\cdot\mathcal{V}_{p}=\partial p^{\alpha}/\partial x^{\alpha}+0=0$;
for $\mathcal{V}_{K}$, considering that $K_{\ \beta}^{\alpha}$ is a similar matrix of
$\mathcal{K}_{\ \beta}^{\alpha}$, $\nabla_{\mathcal{Z}}\cdot\mathcal{V}_{K}=0+\tilde{q}\partial\left(K_{\ \beta}^{\alpha}p^{\beta}\right)/\partial p^{\alpha}=\tilde{q}K_{\ \alpha}^{\alpha}=\tilde{q}\mathcal{K}_{\ \alpha}^{\alpha}=0$;
for $\mathcal{V}_{R}$, $R_{\ \beta}^{\alpha}$ and $\mathcal{R}_{\ \beta}^{\alpha}$
are also similar matrices, thus $\nabla_{\mathcal{Z}}\cdot\mathcal{V}_{R}=0+\tilde{q}\partial\left(R_{\ \beta}^{\alpha}p^{\beta}\right)/\partial p^{\alpha}=\tilde{q}R_{\ \alpha}^{\alpha}=\tilde{q}\mathcal{R}_{\ \alpha}^{\alpha}=0$.
Therefore, after splitting, we get three volume-preserving subsystems,
\begin{eqnarray}
s_{p}^{\tau} & \coloneqq & \begin{cases}
\frac{\mathrm{d}x^{\alpha}}{\mathrm{d}\tau} & =p^{\alpha}\,,\\
\frac{\mathrm{d}p^{\alpha}}{\mathrm{d}\tau} & =0\,,
\end{cases}\label{eq:S_x}\\
s_{K}^{\tau} & \coloneqq & \begin{cases}
\frac{\mathrm{d}x^{\alpha}}{\mathrm{d}\tau} & =0\,,\\
\frac{\mathrm{d}p^{\alpha}}{\mathrm{d}\tau} & =\tilde{q}K_{\ \beta}^{\alpha}p^{\beta}\,,
\end{cases}\label{eq:S_K}\\
s_{R}^{\tau} & \coloneqq & \begin{cases}
\frac{\mathrm{d}x^{\alpha}}{\mathrm{d}\tau} & =0\,,\\
\frac{\mathrm{d}p^{\alpha}}{\mathrm{d}\tau} & =\tilde{q}R_{\ \beta}^{\alpha}p^{\beta}\,.
\end{cases}\label{eq:S_R}
\end{eqnarray}

Now we prove that $s_{p}^{\tau}$, $s_{K}^{\tau}$, and $s_{R}^{\tau}$
are Lorentz invariant systems. Because proofs of the three subsystems
are similar, here we only give the proof of $s_{K}^{\tau}$. We need to derive
the expression of $s_{K}^{\tau}$ in another Lorentz frame $\mathcal{O}'$. For the equation
of $x^{\alpha}$, left-multiply $M$ on both side gives $\mathrm{d}x'^{\alpha}/\mathrm{d}\tau=0$.
For equation of $p^{\alpha}$, left-multiply $M$ on left side gives
$\mathrm{d}p'^{\alpha}/\mathrm{d}\tau$, and on right side
gives $\tilde{q}MK_{\ \beta}^{\alpha}p^{\beta}=\tilde{q}ML\mathcal{K}_{\ \beta}^{\alpha}L^{-1}M^{-1}Mp^{\beta}=\tilde{q}N\mathcal{K}_{\ \beta}^{\alpha}N^{-1}p'^{\beta}=\tilde{q}K_{\ \beta}'^{\alpha}p'^{\beta}$.
The resulting expression of $s_{K}^{\tau}$ in frame $\mathcal{O}'$
is
\begin{eqnarray}
s_{K}'^{\tau} & \coloneqq & \begin{cases}
\frac{\mathrm{d}x'^{\alpha}}{\mathrm{d}\tau} & =0\,,\\
\frac{\mathrm{d}p'^{\alpha}}{\mathrm{d}\tau} & =\tilde{q}K_{\ \beta}'^{\alpha}p'^{\beta}\,,
\end{cases}\label{eq:S_Kprime}
\end{eqnarray}
which has the same form as Eq.\,\ref{eq:S_K}. Therefore, $s_{K}^{\tau}$
is Lorentz invariant.

Using the properties of $K^\alpha_{\ \beta}$ and $R^\alpha_{\ \beta}$, the analytical solutions of $s_{p}^{\tau}$, $s_{K}^{\tau}$, and
$s_{R}^{\tau}$ can be directed obtained as
\begin{eqnarray}
\phi_{p}^{\tau} & \coloneqq & \begin{cases}
x^{\alpha}\left(\tau\right) & =\tau p^{\alpha}+x^{\alpha}\left(0\right)\,,\\
p^{\alpha}\left(\tau\right) & =p^{\alpha}\left(0\right)\,,
\end{cases}\label{eq:phi_x}\\
\phi_{K}^{\tau} & \coloneqq & \begin{cases}
x^{\alpha}\left(\tau\right) & =x^{\alpha}\left(0\right)\,,\\
p^{\alpha}\left(\tau\right) & =\exp\left(\tau\tilde{q}K_{\ \beta}^{\alpha}\right)p^{\beta}\left(0\right)\\
 & =\left[I_{\ \beta}^{\alpha}+\frac{\sinh\left(\tau\tilde{q}E^{r}\right)}{E^{r}}K_{\ \beta}^{\alpha}+\frac{\cosh\left(\tau\tilde{q}E^{r}\right)-1}{\left(E^{r}\right)^{2}}S_{\ \beta}^{\alpha}\right]p^{\beta}\left(0\right)\,,
\end{cases}\label{eq:phi_K}\\
\phi_{R}^{\tau} & \coloneqq & \begin{cases}
x^{\alpha}\left(\tau\right) & =x^{\alpha}\left(0\right)\,,\\
p^{\alpha}\left(\tau\right) & =\exp\left(\tau\tilde{q}R_{\ \beta}^{\alpha}\right)p^{\beta}\left(0\right)\\
 & =\left[I_{\ \beta}^{\alpha}+\frac{\sin\left(\tau\tilde{q}B^{r}\right)}{B^{r}}R_{\ \beta}^{\alpha}+\frac{1-\cos\left(\tau\tilde{q}B^{r}\right)}{\left(B^{r}\right)^{2}}P_{\ \beta}^{\alpha}\right]p^{\beta}\left(0\right)\,,
\end{cases}\label{eq:phi_R}
\end{eqnarray}
where $\exp\left(\cdot\right)$ is the exponential map, $I_{\ \beta}^{\alpha}$
denotes the 4-dimensional identity matrix, and $E^{r}=\sqrt{\left(E_{x}^{r}\right)^{2}+\left(E_{y}^{r}\right)^{2}+\left(E_{z}^{r}\right)^{2}}$
and $B^{r}=\sqrt{\left(B_{x}^{r}\right)^{2}+\left(B_{y}^{r}\right)^{2}+\left(B_{z}^{r}\right)^{2}}$
are respectively the strength of electric and magnetic fields observed in $\mathcal{O}^{r}$. One should notice that the value of $E^{r}$ and $B^{r}$ are evaluated at
$x^{r,\alpha}=L^{-1}x^{\alpha}$. For detailed derivations of Eqs.\,\ref{eq:phi_K}-\ref{eq:phi_R},
please refer to the appendix section.

\subsection{LIVPAs for subsystems}

Based on the exact solutions $\phi_{p}^{\tau}$, $\phi_{K}^{\tau}$, and $\phi_{R}^{\tau}$, the volume-preserving algorithms for corresponding subsystems can be obtained as,
\begin{equation}
\Phi_{p}^{\Delta\tau}\coloneqq\begin{cases}
x^{\alpha,k+1} & =x^{\alpha,k}+\Delta\tau p^{\alpha,k}\,,\\
p^{\alpha,k+1} & =p^{\alpha,k}\,,
\end{cases}\label{eq:PHI_x}
\end{equation}
\begin{equation}
\Phi_{K}^{\Delta\tau}\coloneqq\begin{cases}
x^{\alpha,k+1} & =x^{\alpha,k}\,,\\
p^{\alpha,k+1} & =\left[I_{\ \beta}^{\alpha}+\frac{\sinh\left(\Delta\tau\tilde{q}E^{r,k}\right)}{E^{r,k}}K_{\ \beta}^{\alpha,k}+\frac{\cosh\left(\Delta\tau\tilde{q}E^{r,k}\right)-1}{\left(E^{r,k}\right)^{2}}S_{\ \beta}^{\alpha,k}\right]p^{\beta,k}\,,
\end{cases}\label{eq:PHI_K}
\end{equation}
\begin{equation}
\Phi_{R}^{\Delta\tau}\coloneqq\begin{cases}
x^{\alpha,k+1} & =x^{\alpha,k}\,,\\
p^{\alpha,k+1} & =\left[I_{\ \beta}^{\alpha}+\frac{\sin\left(\Delta\tau\tilde{q}B^{r,k}\right)}{B^{r,k}}R_{\ \beta}^{\alpha,k}+\frac{1-\cos\left(\Delta\tau\tilde{q}B^{r,k}\right)}{\left(B^{r,k}\right)^{2}}P_{\ \beta}^{\alpha,k}\right]p^{\beta,k}\,,
\end{cases}\label{eq:PHI_R}
\end{equation}
where $K_{\ \beta}^{\alpha,k}$, $S_{\ \beta}^{\alpha,k}$,
$R_{\ \beta}^{\alpha,k}$, and $P_{\ \beta}^{\alpha,k}$ denote the
value of $K_{\ \beta}^{\alpha}$, $S_{\ \beta}^{\alpha}$, $R_{\ \beta}^{\alpha}$,
and $P_{\ \beta}^{\alpha}$ evaluated at $x^{\alpha,k}$. $E^{r,k}$
and $B^{r,k}$ are the value of $E^{r}$ and $B^{r}$ calculated at $x^{r,\alpha,k}=L^{-1}x^{\alpha,k}$.

Meanwhile, the implicit midpoint scheme for subsystem $s^\tau_R$ can be proven to be volume-preserving \cite{Ruili_VPA_2015}. The difference equation is
\begin{eqnarray}
\Phi_{Rc}^{\Delta\tau} & \coloneqq & \begin{cases}
x^{\alpha,k+1} & =x^{\alpha,k}\,,\\
p^{\alpha,k+1} & =\left[I_{\ \beta}^{\alpha}-\frac{\Delta\tau}{2}\tilde{q}R_{\ \beta}^{\alpha}\right]^{-1}\left[I_{\ \beta}^{\alpha}+\frac{\Delta\tau}{2}\tilde{q}R_{\ \beta}^{\alpha}\right]p^{\beta,k}\\
 & =\left[I_{\ \beta}^{\alpha}+\frac{2a}{1+a^{2}\left(B^{r}\right)^{2}}R_{\ \beta}^{\alpha}+\frac{2a^{2}}{1+a^{2}\left(B^{r}\right)^{2}}R_{\ \beta}^{\alpha}\right]p^{\beta,k}\,,
\end{cases}\label{eq:Phi_Rc}
\end{eqnarray}
where $a=\Delta\tau\tilde{q}/2$ and the detailed derivations of the compact form are shown in appendix.
The implicit midpoint scheme is equivalent to the Cayley transformation. For a matrix $A$, its Cayley transformation is defined as $\mathrm{cay}\left(A\right)=\left(I-A/2\right)^{-1}\left(I+A/2\right)$, which is the 2-order approximation of $\mathrm{exp}\left(A\right)$. If $A$ is antisymmetric, then the determinant of $\mathrm{cay}\left(A\right)$ equals $1$ \cite{Geometric_numerical_integration}. One should notice that $\Delta\tau\tilde{q}\mathcal{R}_{\ \beta}^{\alpha}$
is an antisymmetric matrix, which thus means $\mathrm{det}\left[\mathrm{cay}\left(\Delta\tau\tilde{q}\mathcal{R}^{\alpha}_{\ \beta}\right)\right]\equiv 1$. Because $\Delta\tau\tilde{q}R_{\ \beta}^{\alpha}=\Delta\tau\tilde{q}L\mathcal{R}_{\ \beta}^{\alpha}L^{-1}$ is the similar matrix of $\Delta\tau\tilde{q}\mathcal{R}_{\ \beta}^{\alpha}$, it is readily to see that $\mathrm{det}\left[\mathrm{cay}\left(\Delta\tau\tilde{q}R^{\alpha}_{\ \beta}\right)\right]\equiv 1$.
Therefore, considering that $\partial p^{\alpha,k+1}/\partial p^{\beta,k}=\mathrm{cay}\left[\Delta\tau\tilde{q}R_{\beta}^{\alpha}\right]$, we have $\left|\partial p^{\alpha,k+1}/\partial p^{\beta,k}\right|\equiv 1$. Consequently, the algorithm is volume-preserving \cite{Ruili_VPA_2015}.

It can be proven that $\Phi_{p}^{\Delta\tau}$, $\Phi_{K}^{\Delta\tau}$,
$\Phi_{R}^{\Delta\tau}$, and $\Phi_{Rc}^{\Delta\tau}$ are all Lorentz
invariant. Similar to the discussion in Sec.\,\ref{sec:howsplitbreaklc}, we need to check the
expressions of these difference equations after transformed into another
Lorentz frame $\mathcal{O}'$. Here, as an example, we only give
the proof of $\Phi_{K}^{\Delta\tau}$. For the difference equation
$x^{\alpha,k+1}=x^{\alpha,k}$, left-multiply $M$ on both sides gives
$x'^{\alpha,k+1}=x'^{\alpha,k}$. For the difference equation of $p^{\alpha}$
in $\Phi_{K}^{\Delta\tau}$, it is of compact form which origins from
$p^{\alpha,k+1}=\exp\left(\Delta\tau\tilde{q}K_{\ \beta}^{\alpha}\right)p^{\beta,k}$.
Left-multiply $M$ on both sides of this exponential difference equation
gives,
\begin{eqnarray}
p'^{\alpha,k+1} & = & M\exp\left(\Delta\tau\tilde{q}K_{\ \beta}^{\alpha}\right)M^{-1}p'^{\beta,k}\\
& = & \exp\left(\Delta\tau\tilde{q}MK_{\ \beta}^{\alpha}M^{-1}\right)p'^{\beta,k} =  \exp\left(\Delta\tau\tilde{q}K_{\beta}'^{\alpha}\right)p'^{\beta,k}\nonumber\\
& = & \left[I_{\ \beta}^{\alpha}+\frac{\sinh\left(\Delta\tau\tilde{q}E^{r,k}\right)}{E^{r,k}}K_{\beta}'^{\alpha}+\frac{\cosh\left(\Delta\tau\tilde{q}E^{r,k}\right)-1}{\left(E^{r,k}\right)^{2}}S_{\beta}'^{\alpha}\right]p'^{\beta,k}\,,\nonumber
\end{eqnarray}
where $E^{r,k}$ and $B^{r,k}$ are the value of $E^{r}$ and
$B^{r}$ at $x^{r,\alpha,k}=N^{-1}x'^{\alpha,k}=L^{-1}x^{\alpha,k}$.
The simplification process of the exponential map is the same as Eq.\,\ref{eq:ProveK}
in the appendix section. Therefore, after Lorentz transformation,
in $\mathcal{O}'$, $\Phi_{K}^{\Delta\tau}$, becomes
\begin{equation}
\Phi_{K}'^{\Delta\tau}\coloneqq\begin{cases}
x'^{\alpha,k+1} & =x'^{\alpha,k}\,,\\
p'^{\alpha,k+1} & = \exp\left(\Delta\tau\tilde{q}K_{\beta}'^{\alpha}\right)p'^{\beta,k}\\
   & =\left[I_{\ \beta}^{\alpha}+\frac{\sinh\left(\Delta\tau\tilde{q}E^{r,k}\right)}{E^{r,k}}K_{\beta}'^{\alpha,k}+\frac{\cosh\left(\Delta\tau\tilde{q}E^{r,k}\right)-1}{\left(E^{r,k}\right)^{2}}S_{\beta}'^{\alpha,k}\right]p'^{\beta,k}\,,
\end{cases}\label{eq:PHI_K-1}
\end{equation}
which has the same form as Eq.\,\ref{eq:PHI_K}. It can also be noticed
that, in an arbitrary Lorentz frame, the value of $E^{r,k}$ is defined
and evaluated in the SRF $\mathcal{O}^{r}$. $E^{r,k}$ works as a
link that connecting the difference equations in different Lorentz frames, which does not break the Lorentz invariance. This can also reflect the reason for choosing a SRF. Similarly, $\Phi_{p}^{\Delta\tau}$, $\Phi_{R}^{\Delta\tau}$, and $\Phi_{Rc}^{\Delta\tau}$ can also be
proven to be Lorentz invariant.

\subsection{High order LIVPAs}

Considering that for Lorentz invariant volume-preserving sub-maps,
composite algorithms of them are also LIVPAs. Therefore, we can compose
them to obtain explicit algorithms with different orders \cite{HeYang_Spliting_2015}. The 1-order
LIVPA can be obtained as
\begin{equation}
\Phi_{1}^{\Delta\tau}=\Phi_{R}^{\Delta\tau}\circ\Phi_{K}^{\Delta\tau}\circ\Phi_{p}^{\Delta\tau}\,.\label{eq:PHI_1}
\end{equation}
Based on the symmetric composite, the 2-order symmetric scheme can
be built as follows,
\begin{equation}
\Phi_{2}^{\Delta\tau}=\Phi_{p}^{\frac{\Delta\tau}{2}}\circ\Phi_{K}^{\frac{\Delta\tau}{2}}\circ\Phi_{R}^{\Delta\tau}\circ\Phi_{K}^{\frac{\Delta\tau}{2}}\circ\Phi_{p}^{\frac{\Delta\tau}{2}}\,.\label{eq:PHI_2}
\end{equation}
For higher order schemes, the $2\left(l+1\right)$-order schemes can
be obtain by \cite{HeYang_Spliting_2015}
\begin{equation}
\Phi_{2\left(l+1\right)}^{\Delta\tau}=\Phi_{2}^{u_{l}\Delta\tau}\circ\Phi_{2}^{w_{l}\Delta\tau}\circ\Phi_{2}^{u_{l}\Delta\tau}\,,\label{eq:PHI_l}
\end{equation}
where $u_{l}=\left(2-2^{1/\left(2l+1\right)}\right)^{-1}$ and $w_{l}=1-2u<0$.
For example, we can get the 4-order LIVPA as
\begin{equation}
\Phi_{4}^{\Delta\tau}=\Phi_{2}^{u_{1}\Delta\tau}\circ\Phi_{2}^{w_{1}\Delta\tau}\circ\Phi_{2}^{u_{1}\Delta\tau}\,,\label{eq:PHI_4}
\end{equation}
where $u_{1}=\left(2-\sqrt[3]{2}\right)^{-1}$and $w_{1}=-\sqrt[3]{2}/\left(2-\sqrt[3]{2}\right)$.
Notice that we can also replace $\Phi_{R}^{\Delta\tau}$ in Eqs.\,\ref{eq:PHI_1},
\ref{eq:PHI_2}, and \ref{eq:PHI_4} by $\Phi_{Rc}^{\Delta\tau}$,
and new types of LIVPAs can be obtained as
\begin{eqnarray}
\Phi_{1c}^{\Delta\tau} & = & \Phi_{Rc}^{\Delta\tau}\circ\Phi_{K}^{\Delta\tau}\circ\Phi_{p}^{\Delta\tau}\,,\label{eq:PHI_1-cay}\\
\Phi_{2c}^{\Delta\tau} & = & \Phi_{p}^{\frac{\Delta\tau}{2}}\circ\Phi_{K}^{\frac{\Delta\tau}{2}}\circ\Phi_{Rc}^{\Delta\tau}\circ\Phi_{K}^{\frac{\Delta\tau}{2}}\circ\Phi_{p}^{\frac{\Delta\tau}{2}}\,,\label{eq:PHI_2-cay}\\
\Phi_{4c}^{\Delta\tau} & = & \Phi_{2c}^{u_{1}\Delta\tau}\circ\Phi_{2c}^{w_{1}\Delta\tau}\circ\Phi_{2c}^{u_{1}\Delta\tau}\,.\label{eq:PHI_4-cay}
\end{eqnarray}
Because $\Phi_{R}^{\Delta\tau}$ is constructed from the exact solution
while $\Phi_{Rc}^{\Delta\tau}$ is a 2-order approximation solution,
algorithms using $\Phi_{R}^{\Delta\tau}$ are theoretically more accurate
than that using $\Phi_{Rc}^{\Delta\tau}$.

Finally, we summarize the key procedures for using the LIVPAs.
\begin{enumerate}
\item \textbf{Preparation Step}: Choose a SRF $\mathcal{O}^r$ and calculate $\mathcal{F}^\alpha_{\ \beta}$ using $F^\alpha_{\ \beta}$ and $L$, namely, $\mathcal{F}^\alpha_{\ \beta}=L^{-1}F^\alpha_{\ \beta}L$. Get the expressions for $E^r$, $B^r$, $K^\alpha_{\ \beta}$, and $R^\alpha_{\ \beta}$. Note that the SRF should not change after determined.
\item \textbf{Iteration Step 1}: Use $x^{\alpha,k}$ to obtain $K^{\alpha,k}_{\ \beta}$ and $R^{\alpha,k}_{\ \beta}$. Substitute $L^{-1}x^{\alpha,k}$ into expressions of $E^r$ and $B^r$ to get $E^{r,k}$ and $B^{r,k}$.
\item \textbf{Iteration Step 2}: Substitute $K^{\alpha,k}_{\ \beta}$, $R^{\alpha,k}_{\ \beta}$, $E^{r,k}$, and $B^{r,k}$ into the difference equations of LIVPAs to get $x^{\alpha,k+1}$ and $p^{\alpha,k+1}$.
\end{enumerate}

\section{Numerical Experiments\label{sec:NumericalExps}}

In this section, we study the numerical performances of LIVPAs in
a typical static axisymmetric electromagnetic field, namely,
\begin{eqnarray}
{\bf A} & = & B_{0}\frac{R^{2}}{3R_{0}}\hat{{\bf e}}_{\theta}\,,\label{eq:EBcirc_A}\\
\varphi & = & E_{0}\frac{R_{0}^{2}}{R}\,,\label{eq:EBcirc_phi}\\
{\bf B} & = & B_{0}\frac{R}{R_{0}}\hat{{\bf e}}_{z}\,,\label{eq:EBcirc_B}\\
{\bf E} & = & E_{0}\frac{R_{0}^{2}}{R^{2}}\hat{{\bf e}}_{R}\,,\label{eq:EBcirc_E}
\end{eqnarray}
where $R=\sqrt{x^{2}+y^{2}}$, $\theta$, and $z$ are cylindrical
coordinates, $\hat{{\bf e}}_{R}$, $\hat{\mathbf{e}}_{\theta}$, and
$\hat{\mathbf{e}}_{z}$ are the unit vectors of cylindrical coordinates,
$B_{0}$ and $E_{0}$ respectively denote the strength of
magnetic and electric fields, and $R_{0}$ is the space parameter
of the field. In this section, we keep several fundamental parameters
unchanged. Expressed in \textbf{SI}, the field parameters are set
as $B_{0}=1\,\mathrm{T}$, $E_{0}=10\,\mathrm{V/m}$, and $R_{0}=\mathrm{m_{0}c}/\mathrm{e}B_{0}\approx1.69\times10^{-3}\,\mathrm{m}$,
where $\mathrm{e}$ is the unit charge. The charge of the particle
is set as $\tilde{q}=1$.

\subsection{Lorentz invariance}

As discussed in Sec.\,\ref{sec:constructLIVPAs}, through performing
Lorentz transformation on difference equations, we can theoretically
test the Lorentz invariance of algorithms. Meanwhile, according to
the definition of Lorentz invariant algorithms, we can also directly use numerical
results to examine the Lorentz invariance. We use the symbol $\Phi_{\mathcal{A}}$ to express the difference equation of an algorithm $\mathcal{A}$ in frame $\mathcal{O}$. Replacing all variables in $\Phi_{\mathcal{A}}$
by variables observed in another Lorentz frame $\mathcal{O}'$, we
can get the difference equation in $\mathcal{O}'$, denoted by $\Theta_{\mathcal{A}}$, which has the same form as $\mathcal{A}$. The initial condition $\mathcal{Z}^{0}$ in $\mathcal{O}$ corresponds to the initial condition $\mathcal{Z}'^{0}=M\mathcal{Z}^{0}$ in $\mathcal{O}'$ for the same process.
The numerical solutions of $\Phi_{\mathcal{A}}$ and $\Theta_{\mathcal{A}}$ calculated
in $\mathcal{O}$ and $\mathcal{O}'$ are thus denoted by $\Phi_{\mathcal{A}}\mathcal{Z}^{0}$
and $\Theta_{\mathcal{A}}\mathcal{Z}'^{0}$, respectively. To compare the results in different frames, we can transform
$\Theta_{\mathcal{A}}\mathcal{Z}'^{0}$ to frame $\mathcal{O}$, namely,
calculating $\mathcal{T}_{M^{-1}}\circ\Theta_{\mathcal{A}}\mathcal{Z}'^{0}$.
If neglecting machine errors, Lorentz invariant algorithms should satisfy that $\Phi_{\mathcal{A}}\mathcal{Z}^{0}=\mathcal{T}_{M^{-1}}\circ\Theta_{\mathcal{A}}\mathcal{Z}'^{0}$.

During the constructions of LIVPAs, there are three Lorentz frames
involved, $\mathcal{O}^{r}$, $\mathcal{O}$, and $\mathcal{O}'$. Since the choice
of $\mathcal{O}^{r}$ is arbitrary, in this section, we set $\mathcal{O}^{r}=\mathcal{O}$,
which means $L=I_{\ \beta}^{\alpha}$ and $N=M$. The expressions
of electromagnetic field in $\mathcal{O}$ is given by Eqs.\,\ref{eq:EBcirc_A}-\ref{eq:EBcirc_E}.
Meanwhile, without loss of generality, we consider $M$ is a subset
of the Lorentz group, namely, the Lorentz boost matrix,
\begin{equation}
L=\frac{\partial x'{}^{\alpha}}{\partial x^{\beta}}=\left(\begin{array}{cccc}
\Gamma & -\Gamma\beta_{1} & -\Gamma\beta_{2} & -\Gamma\beta_{3}\\
-\Gamma\beta_{1} & 1+\frac{\left(\Gamma-1\right)\beta_{1}^{2}}{\beta^{2}} & \frac{\left(\Gamma-1\right)\beta_{1}\beta_{2}}{\beta^{2}} & \frac{\left(\Gamma-1\right)\beta_{1}\beta_{3}}{\beta^{2}}\\
-\Gamma\beta_{2} & \frac{\left(\Gamma-1\right)\beta_{1}\beta_{2}}{\beta^{2}} & 1+\frac{\left(\Gamma-1\right)\beta_{2}^{2}}{\beta^{2}} & \frac{\left(\Gamma-1\right)\beta_{2}\beta_{3}}{\beta^{2}}\\
-\Gamma\beta_{3} & \frac{\left(\Gamma-1\right)\beta_{1}\beta_{3}}{\beta^{2}} & \frac{\left(\Gamma-1\right)\beta_{2}\beta_{3}}{\beta^{2}} & 1+\frac{\left(\Gamma-1\right)\beta_{3}^{2}}{\beta^{2}}
\end{array}\right)\,,\label{eq:L}
\end{equation}
where $\beta=\left|\bm{\beta}\right|$, and $\Gamma=1/\sqrt{1-\beta^{2}}$
is the Lorentz factor. The initial condition of the charged particle
in rest reference frame $\mathcal{O}$ is set as $\mathcal{Z}^{0}=\left(x^{\alpha,0},p^{\alpha,0}\right)$,
where $x^{\alpha,0}=\left(0,0,2,0\right)$ and $p^{\alpha,0}=\left(\sqrt{2},0,1,0\right)$.
We set the speed of frame $\mathcal{O}'$ relative to $\mathcal{O}$
as $\bm{\beta}_{cor}=\left(0.5,0,0\right)$ . Initially, the local
time of the $\mathcal{O}$ and $\mathcal{O}'$ are both set as $0$,
and the origin points of $\mathcal{O}$ and $\mathcal{O}'$ coincide
in 4-spacetime.

\begin{figure}[h]
\centering
\includegraphics[scale=0.6]{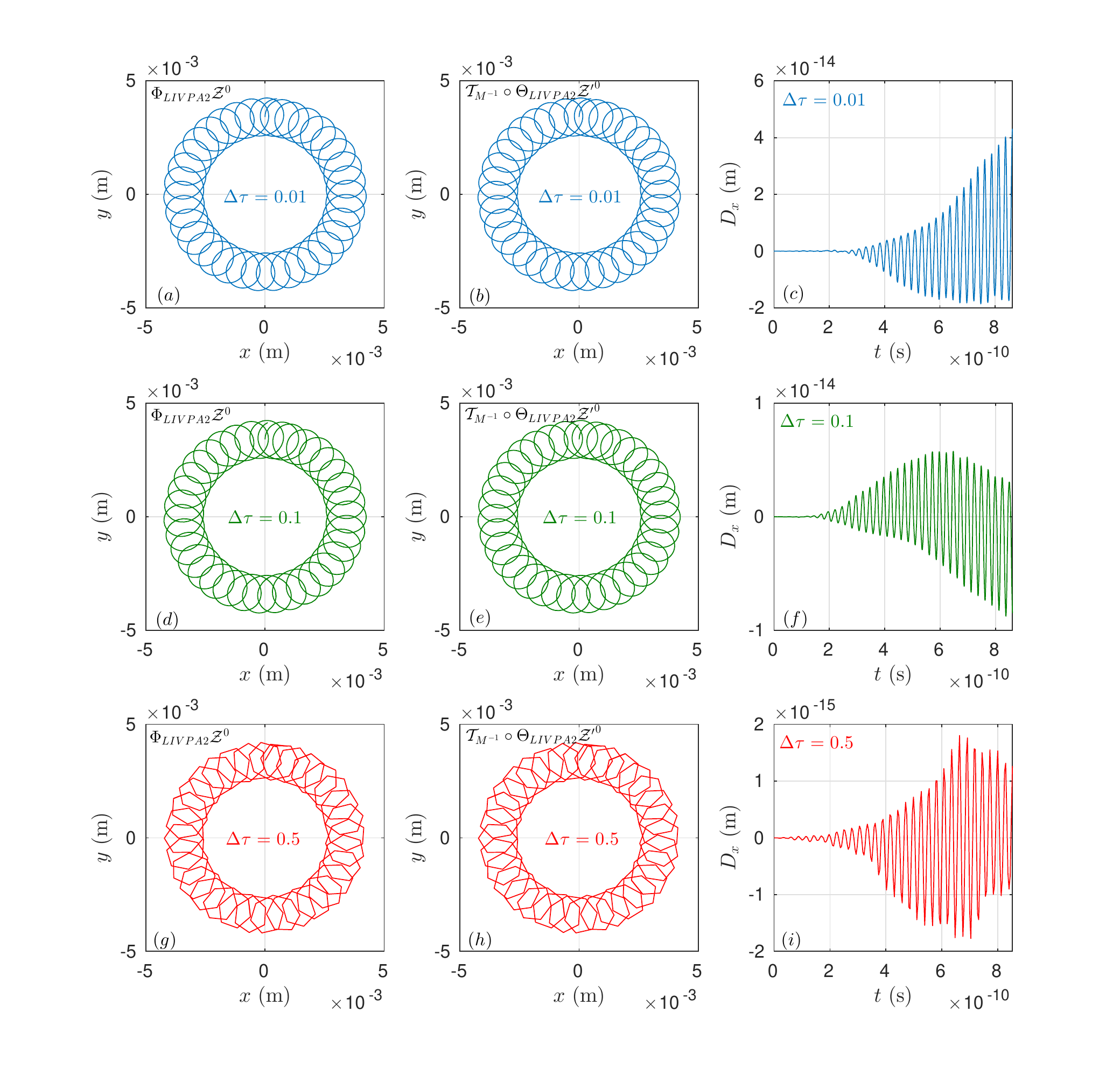}

\caption{Orbits given by 2-order LIVPA $\Phi_{2}$ in different Lorentz frames.
The results in (a), (b) and (c) are simulated with the step-length
$\Delta\tau=0.01$, results in d, e and f are calculated by $\Delta\tau=0.1$,
and results in g, h and i are calculated by $\Delta\tau=0.5$. The
differences between the results calculated in rest and moving frames
are on the order of machine precision.\label{fig:CovTest-LIVPA2}}
\end{figure}

In Figure \ref{fig:CovTest-LIVPA2}, the results of the 2-order LIVPA
$\Phi_{2}$ given by Eq.\,\ref{eq:PHI_2} are depicted. According
to Fig.\,\ref{fig:CovTest-LIVPA2}a and Fig.\,\ref{fig:CovTest-LIVPA2}b,
we can see that the orbit calculated in two frames are consistent.
This can be further explained by Fig.\,\ref{fig:CovTest-LIVPA2}c.
$D_{x}$, the difference of $x$-coordinates in Fig.\,\ref{fig:CovTest-LIVPA2}a
and Fig.\,\ref{fig:CovTest-LIVPA2}b, is on the order of $10^{-14}\,\mathrm{m}$, which
approaches to the machine precision. Meanwhile,
according to Figs.\,\ref{fig:CovTest-LIVPA2}d-\ref{fig:CovTest-LIVPA2}i,
if we increase the step length $\Delta\tau$, though the accuracy
of orbits decreases, the reference-independence of numerical results
still holds. The magnitude of $D_{x}$ for different $\Delta\tau$
keeps unchanged.

\begin{figure}[h]
\centering
\includegraphics[scale=0.7]{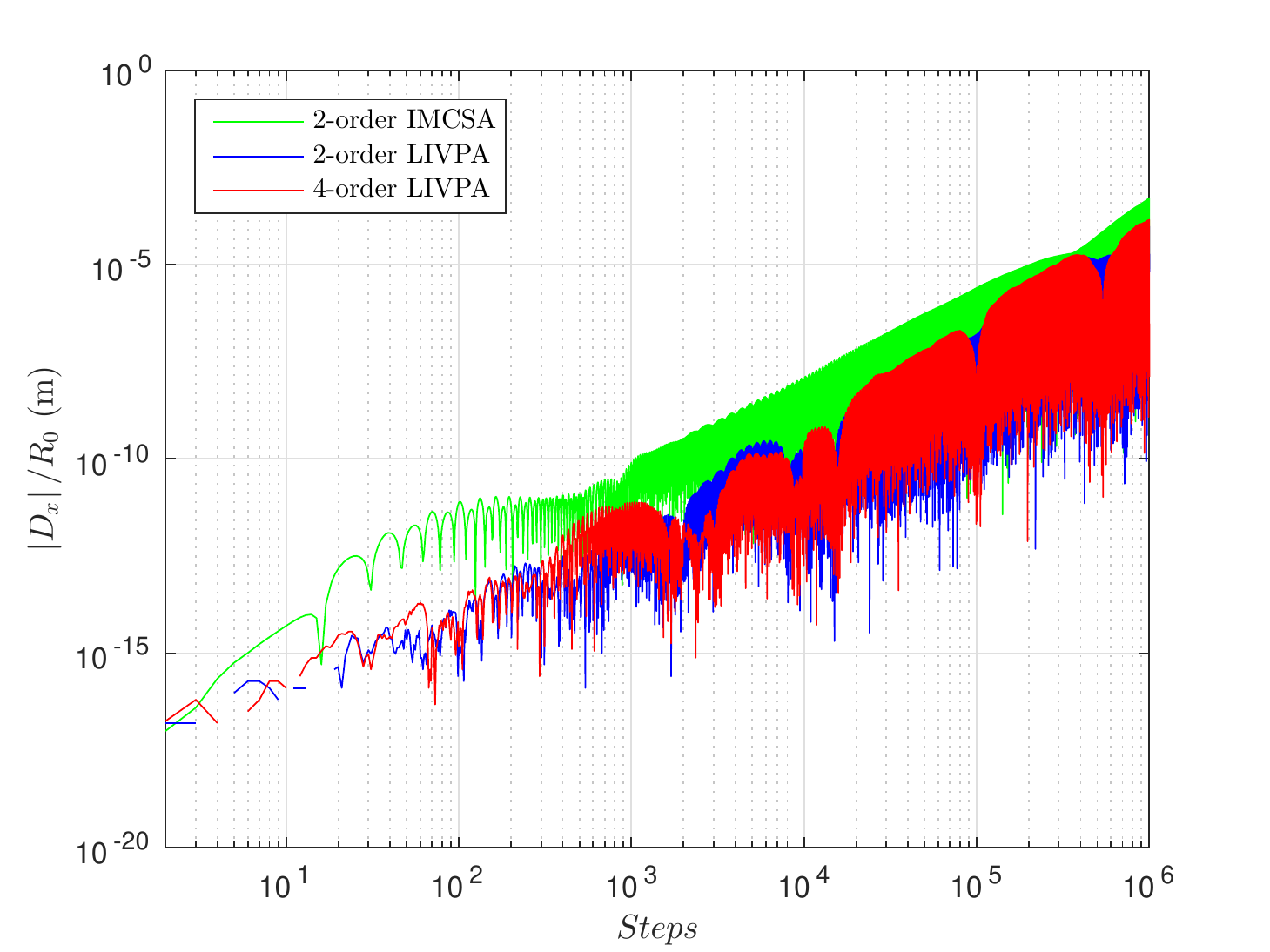}

\caption{Evolution of $\left|D_{x}\right|/R_{0}$ as the increase of iteration
steps. The step-length is set as $\Delta\tau=0.1$. The accumulations
of the 2-order LIVPA (blue line), the 4-order LIVPA (red line), and the implicit midpoint canonical symplectic algorithm (IMCSA, green line) have the same trend. After $10^{6}$ iterations, the order of $D_{x}$
is still ignorable compared with $R_{0}$. The IMCSA, is also a LIA, is obtained by discretizing the 4-dimensional Lorentz invariant Hamiltonian equation of charged particles \cite{Jackson_electrodynamics} using the implicit mid-point symplectic scheme.\label{fig:CovError}}
\end{figure}

In figure \ref{fig:CovTest-LIVPA2}, we plot the first turn of the
orbit and $D_{x}$ is on the order of machine precision. The differences
between the results obtained in different Lorentz inertial frames
by LIVPAs result from the machine truncation errors. As the increase
of iteration times, the accumulation of machine errors cannot be avoided.
Here, we analyze the accumulation of machine errors by studying the
magnitude of $\left|D_{x}\right|/R_{0}$ in terms of iteration step
number, see Fig.\,\ref{fig:CovError}. The step length is set as
$\Delta\tau=0.1$. The same processes are calculated by using 2-order LIVPA, 4-order LIVPA, and the 2-order implicit midpoint canonical symplectic algorithm (IMCSA). The IMCSA is obtained by discretizing the 4-dimensional Lorentz invariant Hamiltonian equation of charged particles \cite{Jackson_electrodynamics} using the implicit mid-point symplectic scheme.
For both cases, the value of $\left|D_{x}\right|/R_{0}$
is on the magnitude of machine error at the beginning and finally
reaches $10^{-5}$ which is still a negligible value after $10^{6}$
iterations.

\begin{figure}[h]
\centering
\includegraphics[scale=0.5]{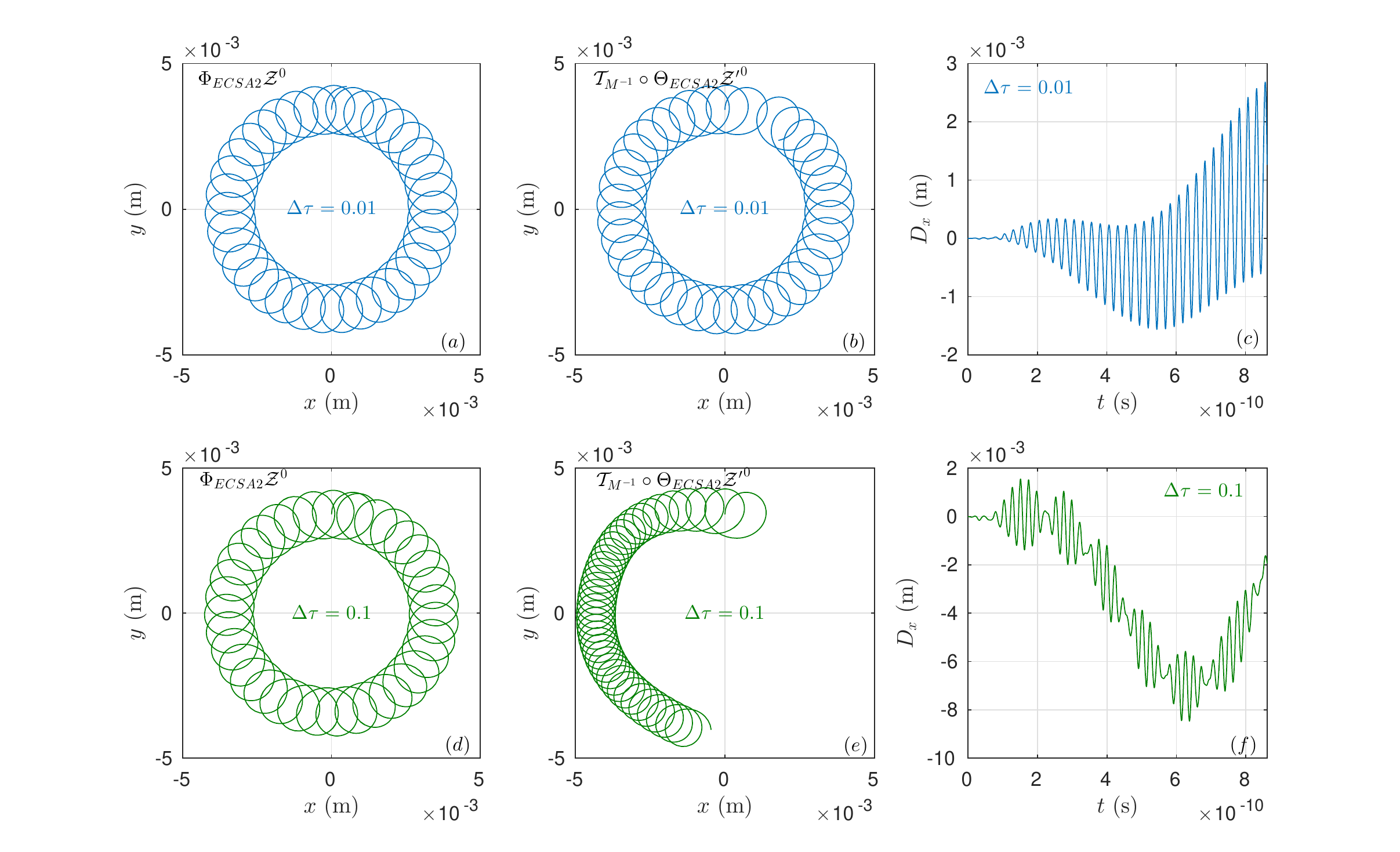}

\caption{Orbits given by 2-order ECSA given in Ref.\,\cite{Zhangruili_2018}
in different Lorentz frames. The results in (a), (b) and (c) are simulated
with the step-length $\Delta\tau=0.01$, while results in d, e and
f are calculated by $\Delta\tau=0.1$. The position difference in
two frames is comparable to $R_{0}$ when $\Delta\tau=0.01$, and
ECSA becomes unstable in the frame $\mathcal{O}'$ for $\Delta\tau=0.1$.
\label{fig:CovTest-ECSA}}
\end{figure}

To compare with LIVPAs, we calculate the same process by use of the
2-order explicit canonical symplectic algorithm (ECSA) given in Ref.\,\cite{Zhangruili_2018}.
This algorithm is built based on generating function method during
which the invariant Hamiltonian is divided into 7 parts. As we have
discussed in Sec.\,\ref{sec:howsplitbreaklc}, the splitting
method can easily break the Lorentz invariance of continuous systems.
In Figure \ref{fig:CovTest-ECSA}, the results in different frames
of $\Phi_{ECSA2}$ are plotted. When we set the step length as $\Delta\tau=0.01$,
the differences of $x$-coordinate calculated by 2-order ECSA in $\mathcal{O}$
and $\mathcal{O}'$ are on the order of $R_{0}$, see Fig.\,\ref{fig:CovTest-ECSA}c.
Especially, as we increase $\Delta\tau$ to $0.1$, ECSA becomes unstable
in $\mathcal{O}'$ and gives incorrect results, see Fig.\,\ref{fig:CovTest-ECSA}e.
Figure \ref{fig:CovTest-ECSA} implies that the 2-order ECSA is not
Lorentz invariant and one should be very careful to use ECSA in moving
frames even though it possesses secular stability in the rest frame.

\begin{figure}[h]
\centering
\includegraphics[scale=0.6]{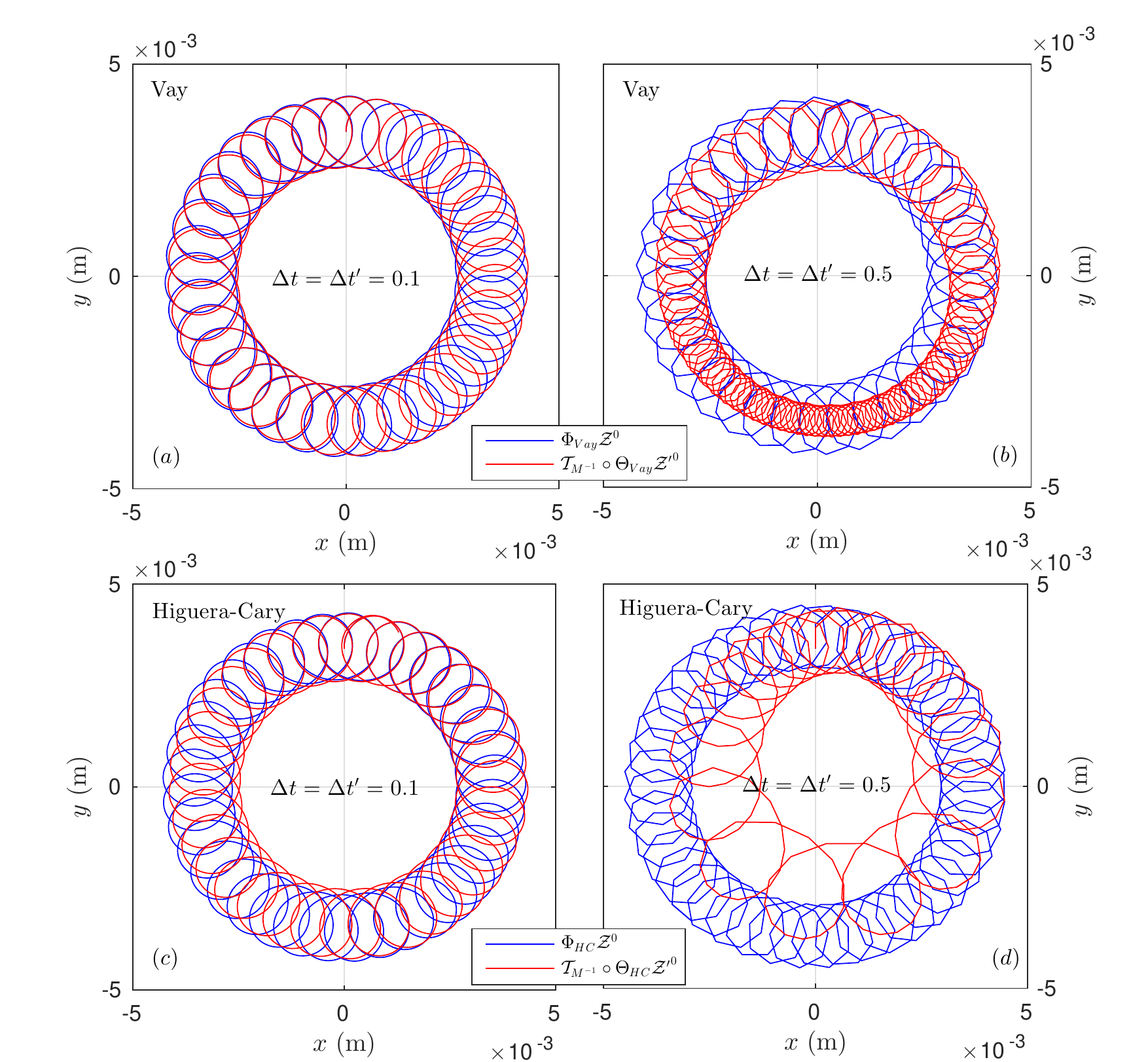}

\caption{The orbits calculated by Vay's method \cite{Vay_2008} and the Higuera-Cary
method \cite{Higuera_Cary_2017}. Results of Vay's method are depicted
in (a) and (b), while results of the Higuera-Cary method are plotted
in (c) and (d). Blue lines give the orbit obtained in frame $\mathcal{O}$
and red lines show the results in frame $\mathcal{O}'$. Both methods
show better reference-independency for small step length. As the step
length increases, the red lines show obvious deviation from blue lines,
which shows that Vay's and Higuera-Cary's method are not Lorentz invariant.
However, both algorithms possess long-term stability in different
step length.\label{fig:VayHC} }
\end{figure}

The Vay scheme \cite{Vay_2008} and the Higuera-Cary scheme \cite{Higuera_Cary_2017} can preserve the
$\mathbf{E}\times\mathbf{B}$ velocity in different Lorentz frames. Both methods show excellent secular
stabilities for simulating relativistic charged particles \cite{Ripperda_2018}. Here,
we use them to solve the same process in Fig.\,\ref{fig:CovTest-LIVPA2}.
In $\mathcal{O}$ and $\mathcal{O}'$, long-term stable orbits can be obtained by both algorithms.
Therefore, in figure \ref{fig:VayHC}, we only depict the first major
turn. The time step is denoted by $\Delta t$ and $\Delta t'$ in
$\mathcal{O}$ and $\mathcal{O}'$, respectively. In the case of $\Delta t=\Delta t'=0.1$,
regardless of the small deviations, the results in different frames are
consistent, see Figs.\,\ref{fig:VayHC}a and c. As the step length increases to $0.5$, however, we can see significant
differences between the red and blue lines. The results show that, Vay's
and Higuera-Cary's method have better Lorentz invariance for small
step length. As the step length increases, even though the global stability still holds, the Lorentz invariance can be broken, which might result from the growth of higher order terms mentioned in Ref.\,\cite{Higuera_Cary_2017}. For LIVPAs, however,
the reference-independent property is not affected by the step length,
see Fig.\,\ref{fig:CovTest-LIVPA2}.

\subsection{Long-term stability}

\begin{figure}[h]
\centering
\includegraphics[scale=0.5]{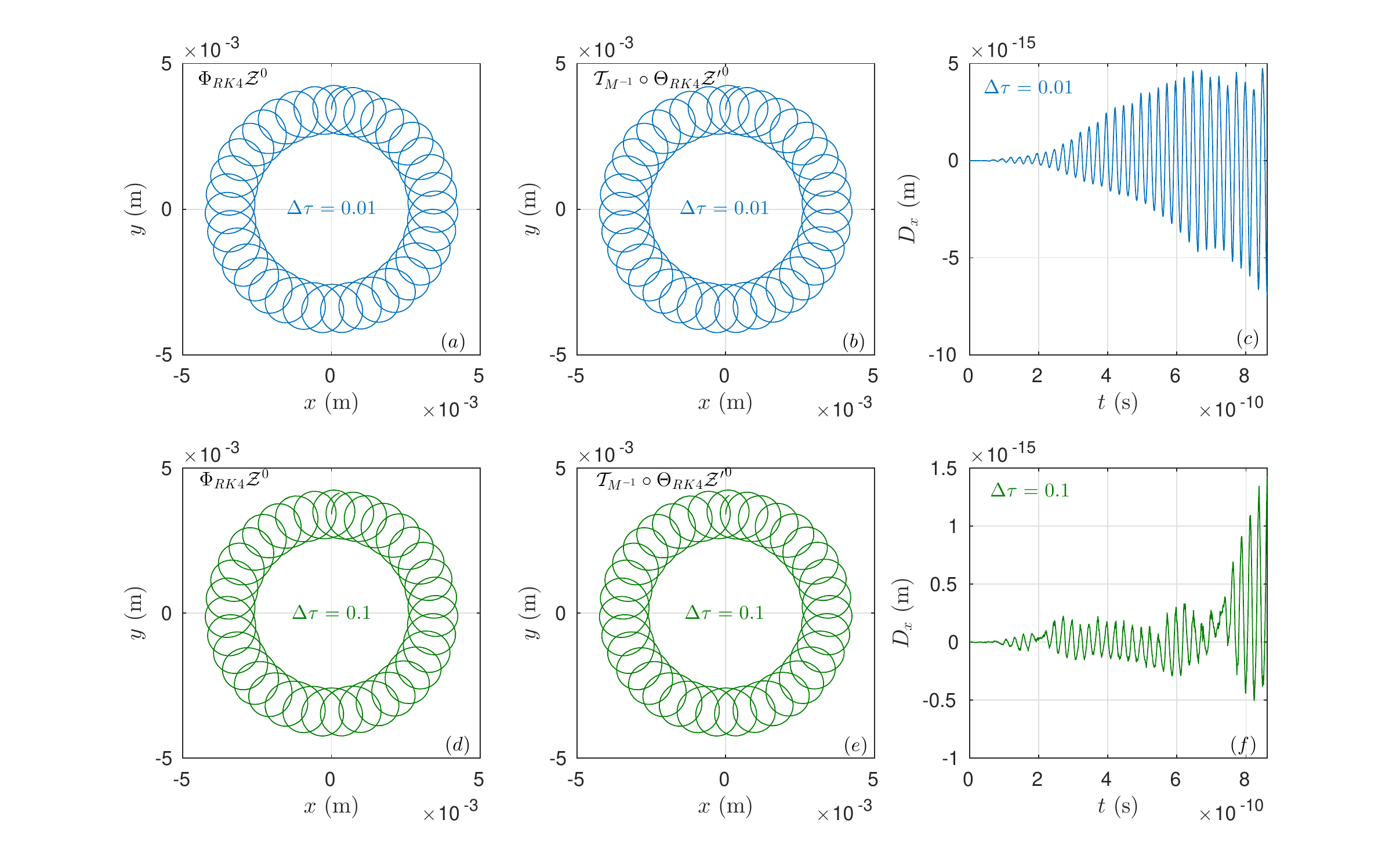}

\caption{Orbits given by 4-order Lorentz invariant Runge-Kutta method in different
Lorentz frames. The results in (a), (b) and (c) are simulated with
the step-length $\Delta\tau=0.01$, while results in d, e and f are
calculated by $\Delta\tau=0.1$. The results show that the 4-order
Runge-Kutta method is Lorentz invariant if it is directly used to
discrete the invariant form of the Lorentz force equation, namely,
Eq.\,\ref{eq:CovLorentzForce_XP}.\label{fig:CovTest-RK4}}
\end{figure}

Through conserving the volume of phase space, the algorithms perform
better in secular simulations compared with traditional algorithms
like the Newton method, the Runge-Kutta method \cite{Qin_Boris_2013,Lee_Qin_PPCF_2015,HeYang_Spliting_2015,Ruili_VPA_2015,Ruili_VPA_CiCP_2016}.
Here we compare the secular stabilities of 2-order LIVPA $\Phi_{2}$
with the Lorentz invariant 4-order Runge-Kutta (RK4) method. The Lorentz
invariant RK4 method is obtained by discretizing Eq.\,\ref{eq:CovLorentzForce_XP}
using the 4-order Runge-Kutta method. Because the discretization keeps
the integrity of all the 4-dimensional geometric objects, the resulting
method is Lorentz invariant, which can also be proven numerically
in figure \ref{fig:CovTest-RK4}.

\begin{figure}[h]
\centering
\includegraphics[scale=0.5]{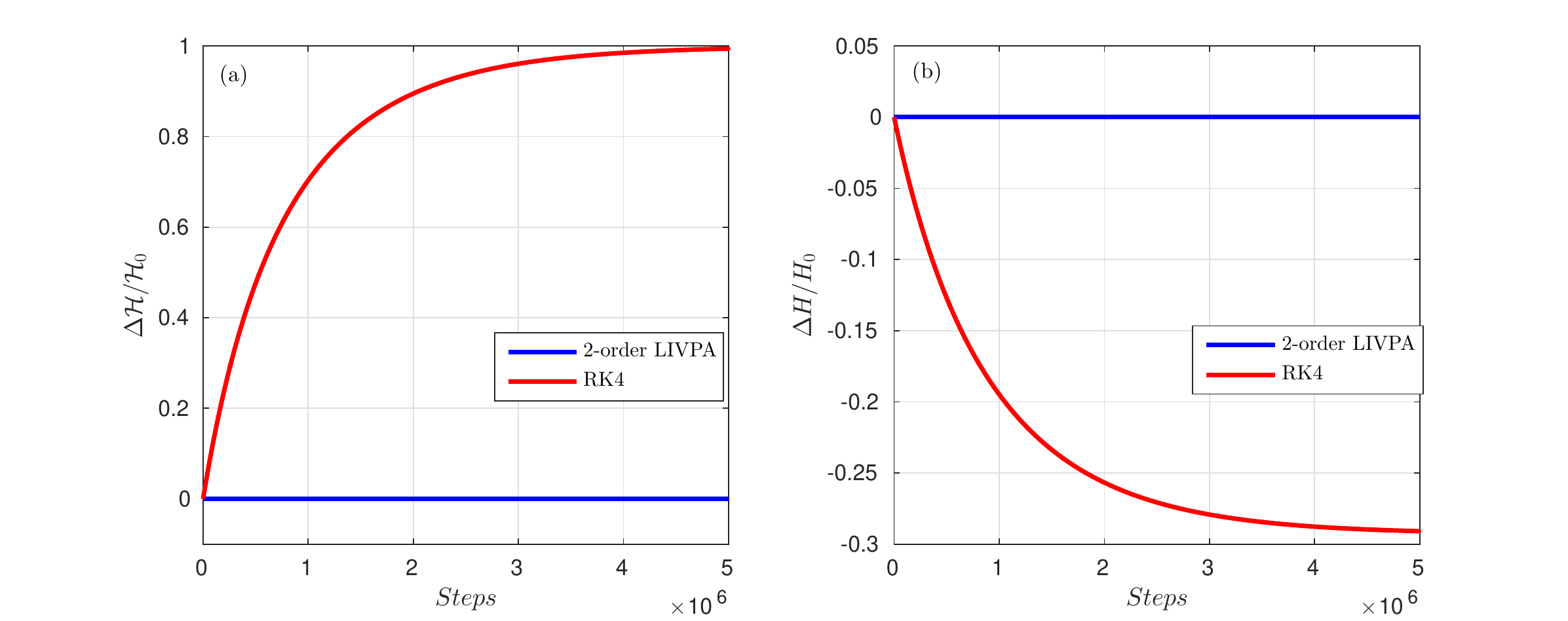}

\caption{Long-term error evolutions of the mass shell (a) and the particle
energy (b) calculated by 2-order LIVPA and 4-order Runge-Kutta. The
results of 2-order LIVPA are depicted by blue lines, while the results
of Runge-Kutta method are plotted by red lines. The total number of
steps is $5\times10^{6}$. The definitions of errors are $\Delta\mathcal{H}=\mathcal{H}\left(n\Delta\tau\right)-\mathcal{H}_{0}$
and $\Delta H=H\left(n\Delta\tau\right)-H_{0}$. The step length is
$\Delta\tau=0.1$.\label{fig:SecularStability}}
\end{figure}

\begin{figure}[h]
\centering
\includegraphics[scale=0.45]{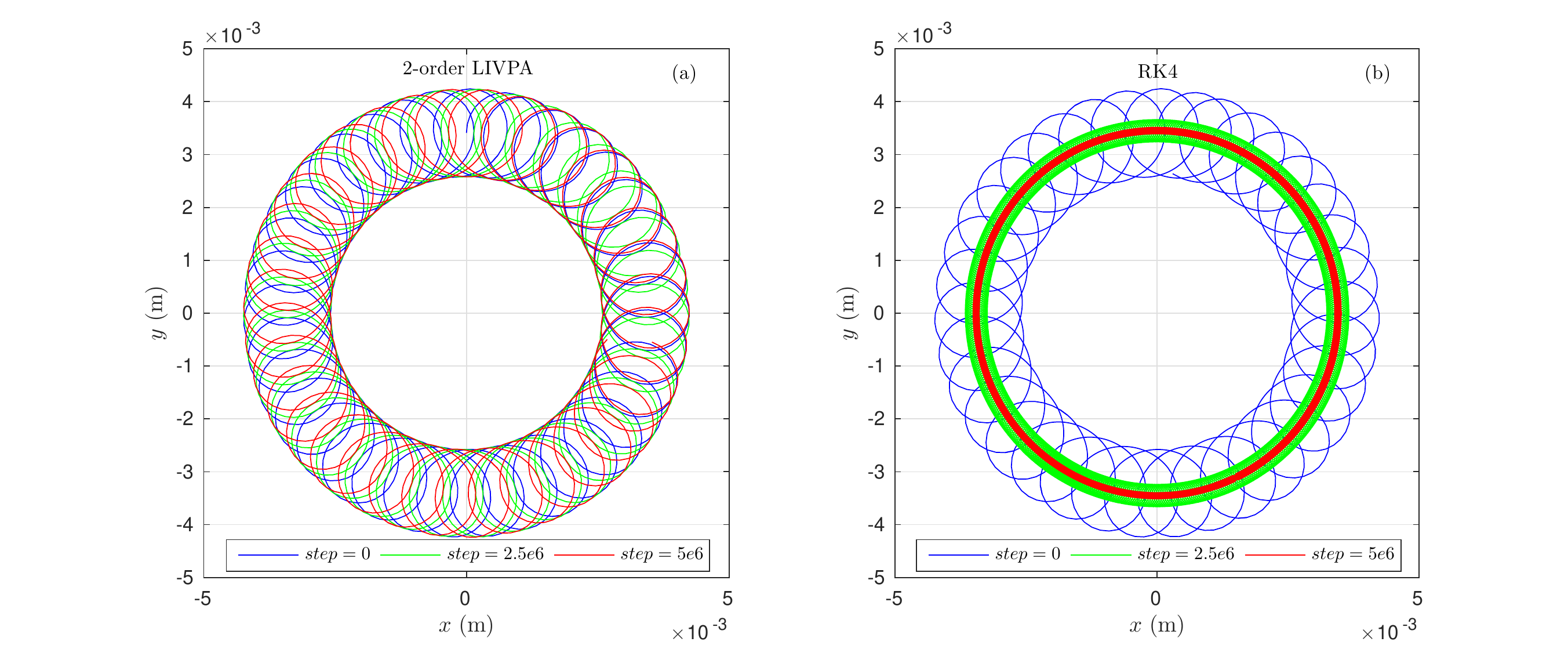}

\caption{Orbits calculated by 2-order LIVPA (a) and 4-order Runge-Kutta (b)
at different moments. The total number of steps is $5\times10^{6}$
and the step length is set as $\Delta\tau=0.1$. \label{fig:Stable_Orbit_circ}}
\end{figure}

For a charged particle moving in static field, there are two important
invariants, namely, the mass-shell
\begin{equation}
\mathcal{H}=g_{\alpha\beta}p^{\alpha}p^{\beta}=\gamma^{2}-{\bf p}^{2}\equiv1\,,\label{eq:DefMassShell}
\end{equation}
and the energy of the particle
\begin{equation}
H=\gamma+\varphi=\sqrt{1+\mathbf{p}^{2}}+\varphi\,,\label{eq:TotalEnergy}
\end{equation}
In figure \ref{fig:SecularStability}, the long-term error evolutions
of the mass shell $\mathcal{H}\left(\mathcal{Z}\right)$ and the particle
energy $H\left(\mathcal{Z}\right)$ simulated by $\Phi_{2}$ and RK4
are depicted. In the case of the mass shell in Fig.\,\ref{fig:SecularStability}a,
the error of mass shell grows significantly to 1 for RK4, while the
error given by 2-order LIVPA is limited near 0. In the case of the
particle energy in Fig.\,\ref{fig:SecularStability}b, the energy
calculated by RK4 decreases 30\% after $5\times10^{6}$ steps, but
the energy obtained by $\Phi_{2}$ conserves well. Therefore, the
long-term stability of 2-order LIVPA is better than RK4, even though
its order is smaller. We also depict the orbits at different moments
obtained by 2-order LIVPA and RK4 in figure \ref{fig:Stable_Orbit_circ}.
The orbits calculated by 2-order LIVPA are depicted in Fig.\,\ref{fig:Stable_Orbit_circ}a,
which remains stable after $5\times10^{6}$ iterations. For RK4, however,
because of the accumulations of numerical errors of RK4, the radius
of rotation on minor period keeps shrinking and incorrect orbits are
obtained after long-term simulation, see Fig.\,\ref{fig:Stable_Orbit_circ}b.

\subsection{Convergence rate of LIVPAs}

\begin{figure}[h]
\centering
\includegraphics[scale=0.7]{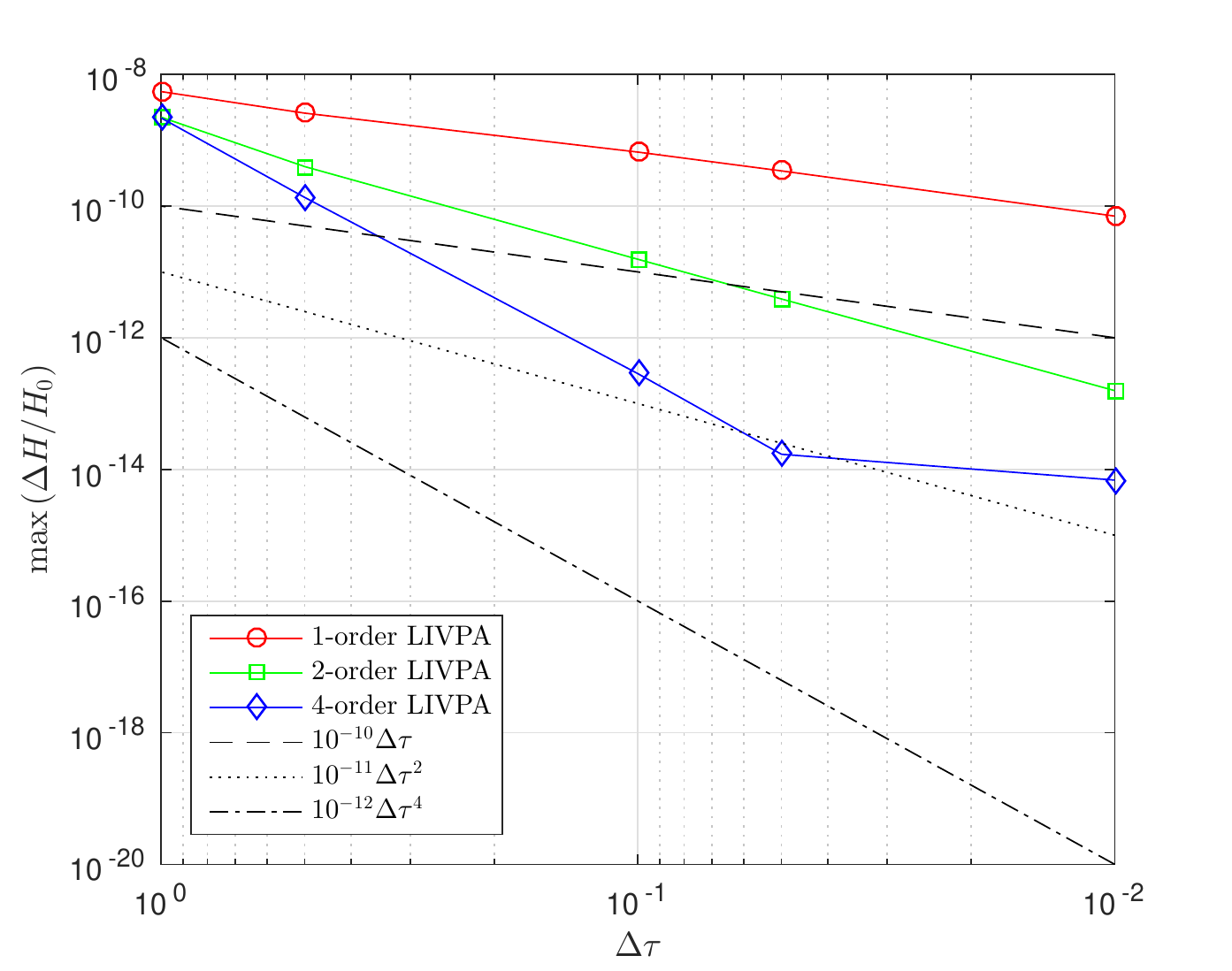}

\caption{Convergence rates of 1-order LIVPA $\Phi_{1}$, 2-order LIVPA $\Phi_{2}$,
and 4-order LIVPA $\Phi_{4}$. The sequences of energy error $\Delta H$
are calculated from $\tau=0$ to $\tau=10$, and the infinity norm
of $\Delta H/H_{0}$ is used to plot the convergence rate. The three
black lines with different slope are reference functions with 1-order,
2-order, and 4-order convergence rates respectively. \label{fig:Orders}}
\end{figure}

In Sec.\,\ref{sec:constructLIVPAs}, we have constructed several
LIVPAs with different orders via different composing procedures. In
order to verify their orders, we perform convergence analysis on the
1-order LIVPA $\Phi_{1}$, 2-order LIVPA $\Phi_{2}$, and 4-order
LIVPA $\Phi_{4}$. The results are depicted in figure \ref{fig:Orders}.
The sequences of energy error $\Delta H$ are calculated from $\tau=0$
to $\tau=10$, and the infinity norm of $\Delta H/H_{0}$ is used
to plot the convergence rate. The red line is the convergence rate
of $\Phi_{1}$. It has the same slope with the function $10^{-10}\Delta\tau$,
which implies $\Phi_{1}$ is a 1-order algorithm. Similarly, $\Phi_{2}$,
plotted by the green line, can be proven to be a 2-order algorithm
via compared with the reference function $10^{-10}\Delta\tau^{2}$.
$\Phi_{4}$ is a 4-order scheme according to the blue line. Because
the value of $\Delta H/H_{0}$ reaches machine errors, the convergence
rate of $\Phi_{4}$ slows down as the step length becomes very small,
see the last point of the blue line at $\Delta\tau=10^{-2}$.

\section{Conclusion\label{sec:Conclusion}}

In this paper, we study the constructions of explicit Lorentz invariant volume-preserving
algorithms for relativistic charged particle dynamics. Through introducing
a splitting reference frame (SRF), we prove that the corresponding splitting operation can avoid breaking the Lorentz invariance of the original system. By use of this procedure, we build explicit LIVPAs with different orders. The Lorentz invariant properties of LIVPAs are tested in a typical electromagnetic field configuration,
which shows that LIVPAs possess reference-independent secular stabilities. It is proven that the Hamiltonian splitting technique for constructing the explicit symplectic algorithms \cite{Zhangruili_2018} breaks the Lorentz invariance. Compared with the Vay scheme \cite{Vay_2008} and the Higuera-Cary scheme \cite{Higuera_Cary_2017}, the benefits of LIVPAs are also obviously reflected. The reference-independency of LIVPAs is not affected by the configurations of step length, while both Vay and Higuera-Cary methods show the decreases of accuracy in different frames when the step length grows.
Meanwhile, long-term behaviors of LIVPAs show better than the Runge-Kutta method in resolving the motion constants such as the mass-shell and the particle energy. Therefore,
LIVPAs have better performances when simulating nonlinear multi-scale
processes. We also provide the numerical convergence analysis to LIVPAs, which proves the ability of the Lorentz invariant splitting method in constructing high order explicit schemes.

The work in this paper extends the method introduced in Ref.\,\cite{Yulei_LCCSA_2016}
in which the integrity of geometric objects is necessary to construct
Lorentz invariant algorithms. We generalize the idea to a more flexible
level, that even though the geometric objects in 4-dimensional spacetime are broken, the Lorentz invariance of algorithms still holds. The combination of
the splitting technique and the Lorentz invariant method provides
a convenient way to build advanced algorithms with high orders.
Additionally, generally speaking, as the orders of algorithms increase, it will be more difficult to implement adaptive time step optimization. Especially, constructing structure-preserving algorithms with adaptive time step is still a difficult task. Some works have been done by use of the time transformation method \cite{Geometric_numerical_integration,Shiyanyan_2019_adaptivestep} which should transform the original system into a new form. The results given by symplectic algorithms of the new system in which the step length are still fixed are equivalent to results by adaptive step length in the original system.
In future work, more works will be done to study and apply the LIVPAs
to key physical problems in different areas, and the usage of LIVPAs in Particle-in-Cell codes will also be studied.

\ack{
This research is supported by Natural Science
Foundation of China (Nos.11805203, 11775222, 11505185), and National Magnetic Confinement Fusion Energy R\&D Program of China (2017YFE0301700).
}

\section*{Appendix}

In this appendix, we provide the detailed derivations of Eqs.\,\ref{eq:phi_K},
\ref{eq:phi_R}, and \ref{eq:Phi_Rc}. According to the definitions
of $\mathcal{K}_{\ \beta}^{\alpha}$ and $\mathcal{R}_{\ \beta}^{\alpha}$,
it is readily to find that $\left(\mathcal{K}_{\ \beta}^{\alpha}\right)^{3}=\left(E^{r}\right)^{2}\mathcal{K}_{\ \beta}^{\alpha}$
and $\left(\mathcal{R}_{\ \beta}^{\alpha}\right)^{3}=-\left(B^{r}\right)^{2}\mathcal{R}_{\ \beta}^{\alpha}$.
Therefore, considering that $E^{r}$ and $B^{r}$ are scalar fields, we have
$\left(K_{\ \beta}^{\alpha}\right)^{3}=L\left(\mathcal{K}_{\ \beta}^{\alpha}\right)^{3}L^{-1}=L\left(E^{r}\right)^{2}\mathcal{K}_{\ \beta}^{\alpha}L^{-1}=\left(E^{r}\right)^{2}L\mathcal{K}_{\ \beta}^{\alpha}L^{-1}=\left(E^{r}\right)^{2}K_{\ \beta}^{\alpha}$.
One should notice that, $E^{r}$ is a function in frame $\mathcal{O}^{r}$
while $K_{\ \beta}^{\alpha}$ is the matrix evaluated in frame $\mathcal{O}$.
When we calculate the value of $\left(K_{\ \beta}^{\alpha}\right)^{3}$
at $x^{\alpha}$ in frame $\mathcal{O}$, the input of value of $E^{r}$
should be $L^{-1}x^{\alpha}$. From the viewpoint of manifold, $x^{\alpha}$
and $L^{-1}x^{\alpha}$ are different coordinates of the same point
on manifold. Similarly, we can also obtain $\left(R_{\ \beta}^{\alpha}\right)^{3}=-\left(B^{r}\right)^{2}R_{\ \beta}^{\alpha}$.

We first prove that $\exp\left(\tau\tilde{q}K_{\ \beta}^{\alpha}\right)=I_{\ \beta}^{\alpha}+\frac{\sinh\left(\tau\tilde{q}E^{r}\right)}{E^{r}}K_{\ \beta}^{\alpha}+\frac{\cosh\left(\tau\tilde{q}E^{r}\right)-1}{\left(E^{r}\right)^{2}}S_{\ \beta}^{\alpha}$
in Eq.\,\ref{eq:phi_K}. Using $\left(K_{\ \beta}^{\alpha}\right)^{2}=S_{\ \beta}^{\alpha}$
and $\left(K_{\ \beta}^{\alpha}\right)^{3}=\left(E^{r}\right)^{2}K_{\ \beta}^{\alpha}$,
we have
\begin{eqnarray}
\exp\left(\tau\tilde{q}K_{\ \beta}^{\alpha}\right) & = & I_{\ \beta}^{\alpha}+\tau\tilde{q}K_{\ \beta}^{\alpha}+\frac{\left(\tau\tilde{q}K_{\ \beta}^{\alpha}\right)^{2}}{2!}+\frac{\left(\tau\tilde{q}K_{\ \beta}^{\alpha}\right)^{3}}{3!}+\cdots\nonumber \\
 & = & I_{\ \beta}^{\alpha}+\tau\tilde{q}K_{\ \beta}^{\alpha}+\frac{\left(\tau\tilde{q}\right)^{2}}{2!}K_{\ \beta}^{\alpha}+\frac{\left(\tau\tilde{q}\right)^{3}\left(E^{r}\right)^{2}}{3!}K_{\ \beta}^{\alpha}+\frac{\left(\tau\tilde{q}\right)^{3}\left(E^{r}\right)^{2}}{4!}S_{\ \beta}^{\alpha}+\cdots\nonumber \\
 & = & I_{\ \beta}^{\alpha}+\frac{1}{E^{r}}\left(\tau\tilde{q}E^{r}+\frac{\left(\tau\tilde{q}E^{r}\right)^{3}}{3!}+\frac{\left(\tau\tilde{q}E^{r}\right)^{5}}{5!}+\cdots\right)K_{\ \beta}^{\alpha}\nonumber \\
 &  & +\frac{1}{\left(E^{r}\right)^{2}}\left(-1+1+\frac{\left(\tau\tilde{q}E^{r}\right)^{2}}{2!}+\frac{\left(\tau\tilde{q}E^{r}\right)^{4}}{4!}+\cdots\right)S_{\ \beta}^{\alpha}\nonumber \\
 & = & I_{\ \beta}^{\alpha}+\frac{1}{E^{r}}\frac{e^{\tau\tilde{q}E^{r}}-e^{-\tau\tilde{q}E^{r}}}{2}K_{\ \beta}^{\alpha}+\frac{1}{\left(E^{r}\right)^{2}}\left(-1+\frac{e^{\tau\tilde{q}E}+e^{-\tau\tilde{q}E}}{2}\right)S_{\ \beta}^{\alpha}\nonumber \\
 & = & I_{\ \beta}^{\alpha}+\frac{\sinh\left(\tau\tilde{q}E^{r}\right)}{E^{r}}K_{\ \beta}^{\alpha}+\frac{\cosh\left(\tau\tilde{q}E^{r}\right)-1}{\left(E^{r}\right)^{2}}S_{\ \beta}^{\alpha}\,.\label{eq:ProveK}
\end{eqnarray}
The derivation of Eqs.\,\ref{eq:phi_R} is similar. The equation
$\exp\left(\tau\tilde{q}R_{\ \beta}^{\alpha}\right)=I_{\ \beta}^{\alpha}+\frac{\sin\left(\tau\tilde{q}B^{r}\right)}{B^{r}}R_{\ \beta}^{\alpha}+\frac{1-\cos\left(\tau\tilde{q}B^{r}\right)}{\left(B^{r}\right)^{2}}P_{\ \beta}^{\alpha}$
can be proven using $\left(R_{\ \beta}^{\alpha}\right)^{2}=P_{\ \beta}^{\alpha}$
and $\left(R_{\ \beta}^{\alpha}\right)^{3}=-\left(B^{r}\right)^{2}R_{\ \beta}^{\alpha}$
as follows,
\begin{eqnarray}
\exp\left(\tau\tilde{q}R_{\ \beta}^{\alpha}\right) & = & I_{\ \beta}^{\alpha}+\tau\tilde{q}R_{\ \beta}^{\alpha}+\frac{\left(\tau\tilde{q}R_{\ \beta}^{\alpha}\right)^{2}}{2!}+\frac{\left(\tau\tilde{q}R_{\ \beta}^{\alpha}\right)^{3}}{3!}+\cdots\nonumber \\
 & = & I_{\ \beta}^{\alpha}+\tau\tilde{q}R_{\ \beta}^{\alpha}+\frac{\left(\tau\tilde{q}\right)^{2}}{2!}P_{\ \beta}^{\alpha}-\frac{\left(\tau\tilde{q}\right)^{3}\left(B^{r}\right)^{2}}{3!}R_{\ \beta}^{\alpha}-\frac{\left(\tau\tilde{q}\right)^{3}\left(B^{r}\right)^{2}}{4!}P_{\ \beta}^{\alpha}+\cdots\nonumber \\
 & = & I_{\ \beta}^{\alpha}+\frac{1}{B^{r}}\left(\tau\tilde{q}B^{r}-\frac{\left(\tau\tilde{q}B^{r}\right)^{3}}{3!}+\frac{\left(\tau\tilde{q}B^{r}\right)^{5}}{5!}-\cdots\right)R_{\ \beta}^{\alpha}\nonumber \\
 &  & -\frac{1}{\left(B^{r}\right)^{2}}\left(-1+1-\frac{\left(\tau\tilde{q}B^{r}\right)^{2}}{2!}+\frac{\left(\tau\tilde{q}B^{r}\right)^{4}}{4!}-\cdots\right)P_{\ \beta}^{\alpha}\nonumber \\
 & = & I_{\ \beta}^{\alpha}+\frac{\sin\left(\tau\tilde{q}B^{r}\right)}{B^{r}}R_{\ \beta}^{\alpha}+\frac{1-\cos\left(\tau\tilde{q}B^{r}\right)}{\left(B^{r}\right)^{2}}P_{\ \beta}^{\alpha}\,.\label{eq:ProveR}
\end{eqnarray}
The explicit form of Cayley transformation in Eq.\,\ref{eq:Phi_Rc}
can be proven as follows,
\begin{eqnarray}
\mathrm{cay}\left(\Delta\tau\tilde{q}R_{\ \beta}^{\alpha}\right) & = & \left(I_{\ \beta}^{\alpha}-aR_{\ \beta}^{\alpha}\right)^{-1}\left(I_{\ \beta}^{\alpha}+aR_{\ \beta}^{\alpha}\right)\nonumber \\
 & = & \left[I_{\ \beta}^{\alpha}+aR_{\ \beta}^{\alpha}+\left(aR_{\ \beta}^{\alpha}\right)^{2}+\left(aR_{\ \beta}^{\alpha}\right)^{3}+\cdots\right]\left(I_{\ \beta}^{\alpha}+aR_{\ \beta}^{\alpha}\right)\nonumber \\
 & = & \left[I_{\ \beta}^{\alpha}+\left(1-\left(aB^{r}\right)^{2}+\left(aB^{r}\right)^{4}-\cdots\right)aR_{\ \beta}^{\alpha}\right.\nonumber \\
 &  & \left.+\left(1-\left(aB^{r}\right)^{2}+\left(aB^{r}\right)^{4}-\cdots\right)a^{2}P_{\ \beta}^{\alpha}\right]\left(I_{\ \beta}^{\alpha}+aR_{\ \beta}^{\alpha}\right)\nonumber \\
 & = & \left[I_{\ \beta}^{\alpha}+\frac{a}{1+\left(aB^{r}\right)^{2}}R_{\ \beta}^{\alpha}+\frac{a^{2}}{1+\left(aB^{r}\right)^{2}}P_{\ \beta}^{\alpha}\right]\left(I_{\ \beta}^{\alpha}+aR_{\ \beta}^{\alpha}\right)\nonumber \\
 & = & \left(I_{\ \beta}^{\alpha}+\frac{2a}{1+\left(aB^{r}\right)^{2}}R_{\ \beta}^{\alpha}+\frac{2a^{2}}{1+\left(aB^{r}\right)^{2}}P_{\ \beta}^{\alpha}\right)\,.\label{eq:ProveRc}
\end{eqnarray}

\section*{References}
\bibliography{Refs}

\begin{thebibliography}{10}
\expandafter\ifx\csname url\endcsname\relax
  \def\url#1{\texttt{#1}}\fi
\expandafter\ifx\csname urlprefix\endcsname\relax\def\urlprefix{URL }\fi
\expandafter\ifx\csname href\endcsname\relax
  \def\href#1#2{#2} \def\path#1{#1}\fi

\bibitem{FengKang_1986}
K.~Feng, Difference schemes for hamiltonian formalism and symplectic geometry,
  Journal of Computational Mathematics 4~(3) (1986) 279--289.

\bibitem{Forest_Ruth_1990}
E.~Forest, R.~D. Ruth, Fourth-order symplectic integration, Physica D 43~(1)
  (1990) 105.

\bibitem{McLachlan_GeoAlgrithm_background}
R.~I. McLachlan, G.~R.~W. Quispel, Geometric integrators for {ODEs}, J. Phys.
  A: Math. Gen. 39~(19) (2006) 5251.

\bibitem{Candy_SympAlg_SepHam_1991}
J.~Candy, W.~Rozmus, A symplectic integration algorithm for separable
  {Hamiltonian} functions, J. Comput. Phys. 92~(1) (1991) 230.

\bibitem{McLachlan_AccuracyOfsymInt1992}
R.~I. McLachlan, P.~Atela, The accuracy of symplectic integrators, Nonlinearity
  5~(2) (1992) 541.

\bibitem{Cary_1993_VMPoison}
J.~R. Cary, I.~Doxas, An explicit symplectic integration scheme for plasma
  simulations, Journal of Computational Physics 107~(1) (1993) 98.

\bibitem{ShangZaijiu_1999}
Z.~Shang, Kam theorem of symplectic algorithms for {Hamiltonian} systems,
  Numerische Mathematik 83~(3) (1999) 477--496.

\bibitem{Qin_VariatianalSymlectic_2008}
H.~Qin, X.~Guan, Variational symplectic integrator for long-time simulations of
  the guiding-center motion of charged particles in general magnetic fields,
  Phys. Rev. Lett. 100~(3) (2008) 035006.

\bibitem{LiJinXing_GC_Symp_2011}
J.~Li, H.~Qin, Z.~Pu, L.~Xie, S.~Fu, Variational symplectic algorithm for
  guiding center dynamics in the inner magnetosphere, Phys. Plasmas 18~(5)
  (2011) 052902.

\bibitem{Kraus_VariationalSym_Thesis}
M.~Kraus, Variational integrators in plasma physics, arXiv preprint
  arXiv:1307.5665.

\bibitem{XiaoJY_PIC_wave_2015}
J.~Xiao, J.~Liu, H.~Qin, Z.~Yu, N.~Xiang, Variational symplectic
  particle-in-cell simulation of nonlinear mode conversion from extraordinary
  waves to {Bernstein} waves, Phys. Plasmas 22~(9) (2015) 092305.

\bibitem{CSPIC_2016}
H.~Qin, J.~Liu, J.~Xiao, R.~Zhang, Y.~He, Y.~Wang, Y.~Sun, J.~W. Burby,
  L.~Ellison, Y.~Zhou, Canonical symplectic particle-in-cell method for
  long-term large-scale simulations of the {Vlasov--Maxwell} equations, Nucl.
  Fusion 56~(1) (2015) 014001.

\bibitem{QiangJi_2017_Sym_VMPoison}
J.~Qiang, Symplectic multiparticle tracking model for self-consistent
  space-charge simulation, Physical Review Accelerators and Beams 20~(1) (2017)
  014203.

\bibitem{Shadwick_Variational_2014}
B.~Shadwick, A.~Stamm, E.~Evstatiev, Variational formulation of macro-particle
  plasma simulation algorithms, Phys. Plasmas 21~(5) (2014) 055708.

\bibitem{Webb_2016_Sym_VMPoison}
S.~D. Webb, A spectral canonical electrostatic algorithm, Plasma Phys.
  Controlled Fusion 58~(3) (2016) 034007.

\bibitem{ZhouYao_2014MHD}
Y.~Zhou, H.~Qin, J.~W. Burby, A.~Bhattacharjee, Variational integration for
  ideal magnetohydrodynamics with built-in advection equations, Phys. Plasmas
  21~(10) (2014) 102109.

\bibitem{ZhouYao_2016_PRE}
Y.~Zhou, Y.-M. Huang, H.~Qin, A.~Bhattacharjee, Formation of current
  singularity in a topologically constrained plasma, Physical Review E 93~(2)
  (2016) 023205.

\bibitem{XiaoJY_2016_NonCanSymTwoFluid}
J.~Xiao, H.~Qin, P.~J. Morrison, J.~Liu, Z.~Yu, R.~Zhang, Y.~He, Explicit
  high-order noncanonical symplectic algorithms for ideal two-fluid systems,
  Phys. Plasmas 23~(11) (2016) 112107.

\bibitem{McLachlan_Symplectic_KDV}
U.~M. Ascher, R.~I. McLachlan, Multisymplectic box schemes and the
  {Korteweg--de Vries} equation, Applied Numerical Mathematics 48~(3-4) (2004)
  255.

\bibitem{QinMengZhao_SympNSL}
J.-Q. Sun, M.-Z. Qin, Multi-symplectic methods for the coupled 1d nonlinear
  {Schr\"{o}dinger} system, Comput. Phys. Commun. 155~(3) (2003) 221.

\bibitem{Geometric_numerical_integration}
E.~Hairer, C.~Lubich, G.~Wanner, Geometric numerical integration:
  structure-preserving algorithms for ordinary differential equations, Vol.~31,
  Springer Science \& Business Media, 2006.

\bibitem{Xiaojy_2018_PST}
J.~Xiao, H.~Qin, J.~Liu, Structure-preserving geometric particle-in-cell
  methods for {Vlasov-Maxwell} systems, Plasma Sci. Technol 20~(11) (2018)
  110501.

\bibitem{ZhangRuili_ExpGenerateSym_2016}
R.~Zhang, H.~Qin, Y.~Tang, J.~Liu, Y.~He, J.~Xiao, Explicit symplectic
  algorithms based on generating functions for charged particle dynamics, Phys.
  Rev. E 94~(1) (2016) 013205.

\bibitem{Zhangruili_2018}
R.~Zhang, Y.~Wang, Y.~He, J.~Xiao, J.~Liu, H.~Qin, Y.~Tang, Explicit symplectic
  algorithms based on generating functions for relativistic charged particle
  dynamics in time-dependent electromagnetic field, Phys. Plasmas 25~(2) (2018)
  022117.

\bibitem{ZhouZhaoQi_2017_ExpSymp}
Z.~Zhou, Y.~He, Y.~Sun, J.~Liu, H.~Qin, Explicit symplectic methods for solving
  charged particle trajectories, Phys. Plasmas 24~(5) (2017) 052507.

\bibitem{HeYang_Ksymp_PLA_2016}
Y.~He, Z.~Zhou, Y.~Sun, J.~Liu, H.~Qin, Explicit {K-symplectic} algorithms for
  charged particle dynamics, Phys. Lett. A 381~(6) (2016) 568--573.

\bibitem{XiaoJY_2019_NonSym}
J.~Xiao, H.~Qin, Explicit high-order gauge-independent symplectic algorithms
  for relativistic charged particle dynamics, Comput. Phys. Commun. 241 (2019)
  19.

\bibitem{Birdsall_Book}
C.~K. Birdsall, A.~B. Langdon, Plasma physics via computer simulation, CRC
  Press, 2004.

\bibitem{PSC_2016}
K.~Germaschewski, W.~Fox, S.~Abbott, N.~Ahmadi, K.~Maynard, L.~Wang, H.~Ruhl,
  A.~Bhattacharjee, The plasma simulation code: A modern particle-in-cell code
  with patch-based load-balancing, Journal of Computational Physics 318 (2016)
  305.

\bibitem{Ripperda_2018}
B.~Ripperda, F.~Bacchini, J.~Teunissen, C.~Xia, O.~Porth, L.~Sironi,
  G.~Lapenta, R.~Keppens, A comprehensive comparison of relativistic particle
  integrators, The Astrophysical Journal Supplement Series 235~(1) (2018) 21.

\bibitem{Higuera_Cary_2017}
A.~V. Higuera, J.~R. Cary, Structure-preserving second-order integration of
  relativistic charged particle trajectories in electromagnetic fields, Phys.
  Plasmas 24~(5) (2017) 052104.

\bibitem{VPA_covLorentz_2017}
A.~Matsuyama, M.~Furukawa, High-order integration scheme for relativistic
  charged particle motion in magnetized plasmas with volume preserving
  properties, Comput. Phys. Commun. 220 (2017) 285.

\bibitem{Qin_Boris_2013}
H.~Qin, S.~Zhang, J.~Xiao, J.~Liu, Y.~Sun, W.~M. Tang, Why is {Boris} algorithm
  so good?, Phys. Plasmas 20~(8) (2013) 084503.

\bibitem{HeYang_Spliting_2015}
Y.~He, Y.~Sun, J.~Liu, H.~Qin, Volume-preserving algorithms for charged
  particle dynamics, J. Comput. Phys. 281 (2015) 135.

\bibitem{Ruili_VPA_2015}
R.~Zhang, J.~Liu, H.~Qin, Y.~Wang, Y.~He, Y.~Sun, Volume-preserving algorithm
  for secular relativistic dynamics of charged particles, Phys. Plasmas 22~(4)
  (2015) 044501.

\bibitem{He_2016_Highorder_RVPA}
Y.~He, Y.~Sun, R.~Zhang, Y.~Wang, J.~Liu, H.~Qin, High order volume-preserving
  algorithms for relativistic charged particles in general electromagnetic
  fields, Phys. Plasmas 23~(9) (2016) 092109.

\bibitem{Vay_PRL_2007_RevEffets}
J.~L. Vay, Noninvariance of space- and time-scale ranges under a lorentz
  transformation and the implications for the study of relativistic
  interactions, Phys. Rev. Lett. 98~(13) (2007) 130405.

\bibitem{Vay_2008}
J.~L. Vay, Simulation of beams or plasmas crossing at relativistic velocity,
  Phys. Plasmas 15~(5) (2008) 056701.

\bibitem{Yulei_LCCSA_2016}
Y.~Wang, J.~Liu, H.~Qin, Lorentz covariant canonical symplectic algorithms for
  dynamics of charged particles, Phys. Plasmas 23~(12) (2016) 122513.

\bibitem{Jackson_electrodynamics}
J.~D. Jackson, Classical electrodynamics, Vol.~3, Wiley New York etc., 1962.

\bibitem{Ruili_VPA_CiCP_2016}
R.~Zhang, J.~Liu, H.~Qin, Y.~Tang, Y.~He, Y.~Wang, Application of lie algebra
  in constructing volume-preserving algorithms for charged particles dynamics,
  Communications in Computational Physics 19~(5) (2016) 1397.

\bibitem{Lee_Qin_PPCF_2015}
C.~L. Ellison, J.~Finn, H.~Qin, W.~M. Tang, Development of variational guiding
  center algorithms for parallel calculations in experimental magnetic
  equilibria, Plasma Phys. Controlled Fusion 57~(5) (2015) 054007.

\bibitem{Shiyanyan_2019_adaptivestep}
Y.~Shi, Y.~Sun, Y.~Wang, J.~Liu, Study of adaptive symplectic methods for
  simulating charged particle dynamics, Journal of Computational Dynamics 6~(2)
  (2019) 429.

\end{thebibliography}

\end{document}